\documentclass[journal]{IEEEtran}
\usepackage[pdftex]{graphicx}
\usepackage{subfigure}
\usepackage{multirow}
\usepackage[cmex10]{amsmath}
\usepackage{tabularx}
\usepackage{breqn}
\usepackage{siunitx}
\usepackage{amsfonts}
\usepackage[mathscr]{euscript}
\usepackage{epstopdf}
\usepackage{amsmath}
\usepackage{amsfonts}
\usepackage{amssymb}
\usepackage{cleveref}
\usepackage{lettrine}
\usepackage{caption}
\newenvironment{definition}[1][Definition]{\begin{trivlist}
\item[\hskip \labelsep {\bfseries #1}]}{\end{trivlist}}

  \newcommand{\bs}{\mathbf{s}}

  \newcommand{\bv}{\mathbf{v}}

  \newcommand{\bx}{\mathbf{x}}

  \newcommand{\bF}{\mathbf{F}}

  \newcommand{\bH}{\mathbf{H}}

  \newcommand{\bR}{\mathbf{R}}

  \newcommand{\bU}{\mathbf{U}}

\newtheorem{lemma}{Lemma}

\newtheorem{theorem}{Theorem}

\newcommand{\beq}{\begin{equation}}

\newcommand{\enq}{\end{equation}}

\newcommand{\beqa}{\begin{eqnarray}}

\newcommand{\enqa}{\end{eqnarray}}

\newcommand{\bea}{\begin{array}}

\newcommand{\ena}{\end{array}}

\newcommand{\bef}{\begin{figure}}

\newcommand{\enf}{\end{figure}}

\newcommand{\bds}{\begin {itemize}}

\newcommand{\eds}{\end {itemize}}

\newcommand{\bdf}{\begin{definition}}

\newcommand{\blm}{\begin{lemma}}

\newcommand{\edf}{\end{definition}}

\newcommand{\elm}{\end{lemma}}

\newcommand{\bthm}{\begin{theorem}}

\newcommand{\ethm}{\end{theorem}}

\newcommand{\cI}{{\ensuremath{\mathcal{I}}}}

\DeclareMathAlphabet{\mathcalligra}{T1}{calligra}{m}{n}

\begin{document}

\title{Diversity Pulse Shaped Transmission in Ultra-Dense Small Cell Networks}

\author{Amir H.~Jafari{$^{1}$}, Vijay Venkateswaran{$^{3}$}, David~L\'opez-P\'erez{$^{2}$}, Jie~Zhang{$^{1}$}\\

\
{$~^{1}$}Dept. of Electronic \& Electrical Engineering, University of Sheffield, Sheffield S1 3JD, UK\\
{$~^{2}$}Bell Laboratories Nokia, Dublin, Ireland\\
{$~^{3}$}Huawei Technologies, Sweden\\}
\maketitle

\maketitle

\begin{abstract}
In ultra-dense small cell networks,
spatial multiplexing gain is a challenge because of the different propagation conditions. 
The channels associated with different transmit-receive pairs can be highly correlated due to the 
i) high probability of line-of-sight (LOS) communication between user equipment (UE) and base station (BS), and 
ii) insufficient spacing between antenna elements at both UE and BS. 
In this paper, we propose a novel transmission technique titled Diversity Pulse Shaped Transmission (DPST) to enhance the throughput over the correlated MIMO channels in an ultra-dense small cell network. 
The fundamental of DPST is to shape transmit signals at adjacent antennas with distinct interpolating filters, introducing pulse shaping diversity. 
In DPST, each antenna transmits its own data stream with a relative deterministic time offset-which must be a fraction of the symbol period-with respect to the adjacent antenna. 
The delay is interpolated with the pulse shaped signal generating a virtual MIMO channel that benefits from increased diversity from the receiver perspective. 
To extract the diversity, the receiver must operate in an over-sampled domain and hence a fractionally spaced equaliser (FSE) is proposed. 
The joint impact of DPST and FSE helps the receiver to sense a less correlated channel, eventually enhancing the UE's throughput. 
Moreover, in order to minimise the spatial correlation, we aim to optimise the deterministic fractional delay. 
Simulation results show that applying DPST to a correlated channel can approximately enhance the UE throughput by 1.93x and 3.76x in $2\times2$ and $4\times4$ MIMO systems, respectively.

\end{abstract}

\section{Introduction}

In order to provide the 100$\times$ network capacity increase and meet the traffic demands forecasted for 2020, 
network densification is envisioned as the most promising technology, since it has the potential to increase the network capacity proportionally with the number of deployed cells \cite{2015Lopez}. An Ultra-dense small cell network features dense orthogonal deployment of small cell base stations (BSs) with the macrocell tier, and aims to boost the coverage and capacity of cellular networks through extensive spatial reuse of the spectrum \cite{Eurasip}. According to \cite{fronthaul}, an ultra-dense small cell network is defined as a cellular network with traffic volume per area greater than 700 $ \rm Gbps/km^{2}$ or user equipment (UE) density greater than 0.2 $ \rm UEs/m^{2}$, implying that UE density is smaller than the BS density. Ultra-dense small cell networks aim to provide more capacity through offloading from macrocells in public places with a large number of UEs such as airports and shopping malls as well as indoor environments, where there is a degradation in link margin and throughput due to absorption loss by walls \cite{Nokia_UDN}. 

Multiple input multiple output (MIMO) technology also has the potential to increase the network capacity proportional to the minimum number of transmit/receive antennas~\cite{1532224}. One flavour of MIMO technology is spatial multiplexing, which subject to availability of distinct propagation paths for each transmit/receive pair allows simultaneous transmission of independent data streams from different transmit antennas, eventually enhancing the network capacity by the minimum number of transmit/receive antennas~\cite{LTE}. 
However, it is not clear whether both ultra-dense small cell networks and MIMO technologies can be exploited simultaneously. Indeed, applying spatial multiplexing gain to ultra-dense small cell networks is challenging due to the different propagation conditions when compared to macrocell networks. 
For instance, 
in macrocell networks, spatial multiplexing gain typically improves when the MIMO channel benefits from sufficient environment scattering enabling different channel pairs to be uncorrelated. 
In contrast,
in ultra-dense small cell networks, spatial multiplexing gains may be limited due to channel pairs being correlated. 
Two phenomena contribute to this channel spatial correlation. 
Firstly, due to proximity of both UE and BS, 
there is a high probability of line-of-sight (LOS) communication 
(the channel can even be prone to only LOS communication as the inter-site-distance (ISD) is decreased). 
Secondly, the antennas at both UE and BS may be placed very closely to each other ($\sim$ half wavelength). 
This spatial correlation causes the communication channel to be ill-conditioned, 
which lowers the number of independent parallel data streams that can be simultaneously multiplexed and decoded. 
As a result, MIMO throughput gains are significantly degraded when compared to macrocell networks,
and thus we may be trading spatial reuse for spatial multiplexing when densifying the network. 

\subsubsection*{\textbf{Central Idea}}

In order to compensate for the reduction in spatial multiplexing gain and enhance the UE's throughput over correlated MIMO channels in an ultra-dense small cell network, 
novel transmission schemes have to be investigated. 
In this regard, we propose diversity pulse shaped transmission (DPST) as a novel transmission technique, 
which exploits distinct pulse shapes to modulate data streams of adjacent antennas~\cite{Jafari-SPAWC}. 

DPST is based on the principle that the signals corresponding to adjacent antenna elements should be shaped with distinct band limited pulse shaping filters to enhance diversity. 
The filters are here characterised by \emph{deterministic delays}, 
meaning that the pulse shaped signal at one antenna is delayed by a deterministic time amount with respect to its adjacent one. This leads to inter-symbol interference (ISI) in the time domain,
which in turn generates diversity among different channel pairs. 

It is important to note that 
the deterministic delay must be a \emph{fraction of the symbol period} in order to allow the UE receiver to resolve the multiple delayed replicas of the transmitted data stream within the symbol period (in contrast to overlapping replicas in correlated scenarios), and hence distinguish between the transmissions of different transmit antennas.

For simplicity, we consider a $2\times2$ MIMO setup. 
In this case, the receive antennas would observe transmit antenna~1 to transmit its data stream with symbol period $T_s$ as well as transmit antenna~2 to transmit its data stream with delay $\tau$ with respect to antenna~1, 
while being sampled at $T_s$  where $0 < \tau < T_{s}$. 
This implies that in a LOS scenario and assuming that the receiver is synchronised to $T_s$ (ignoring the bulk delays), 
the receive antennas would observe a direct path from transmit antenna~1 with data stream~1 as well as a delayed path from transmit antenna~2 with data stream~2, with the latter being corrupted by ISI where the generated ISI helps to diversify the different channel pairs. Note that while we restrict our modelling to a $2\times2$ MIMO case, 
DPST can be applied to arbitrary sized antenna arrays.

In order to compensate for the increased ISI in one of the streams, we propose to use a fractionally spaced equaliser (FSE) at the UE, which operates on the precoded data streams and wireless channel output and eventually improves the UE's throughput in ultra-dense small cell networks~\cite{489269}. In order to ensure a reasonable estimate of multi-path signals, minimum mean squared error (MMSE) is used to design the equaliser.

\subsubsection*{\textbf{Contributions}}

In this paper, we present DPST, 
whose combined arrangement based on pulse shaping diversity and deterministic delay as well as oversampled receiver through FSE improves the overall dimensionality of the transmitted multi-antenna data streams viewed by the receiver.  Consequently, DPST is able to compensate for the loss of spatial multiplexing gain in ulta-dense small cell networks and enhances the data rates by almost the minimum number of transmit/receive antennas.  
We further discuss the optimisation of the deterministic fractional delay that minimizes the correlation among channel pairs. Moreover, we evaluate the performance of DPST in a single tier hexagonal small cell layout,
considering downlink communications. In more detail, we quantify the degradation of signal-to-interference-plus-noise ratio (SINR) and UE throughput due to spatial correlation in $2\times2$ and $4\times4$ MIMO systems at different ISDs, i.e., 20~m, 50~m and 100~m, 
and then show that the proposed DPST arrangement will lead to significant gains, 
e.g., for an ISD of 50~m, DPST provides a 50\%-tile SINR improvement of 10.5~dB and 12~dB in $2\times2$ and $4\times4$ MIMO systems, respectively,
and enhance the UE throughput by approximately 2x and 4x in $2\times2$ and $4\times4$ MIMO systems, respectively. We also show that optimum fractional delay that is derived through optimization leads to 1.86x gain in UE throughput with respect to non-optimum delay for an ISD of 50~m.

The rest of this paper is as follows. 
Section~\ref{sec:part_Correlation} provides a channel correlation model for ultra-dense small cell networks. 
Section~\ref{sec:part_DPST} details the proposed DPST, the fractionally spaced receiver and the precoder designs to estimate the transmitted signals.
Moreover, it also discusses the cyclic delay diversity (CDD) transmission technique and explains how it differs from DPST to clarify any misunderstandings. 
Section~\ref{sec:part_Optimization} introduces the optimisation algorithm to derive the optimised delay in order to achieve the least degree of correlation among channel pairs. 
Section~\ref{sec:part_Simulation} presents a performance evaluation/comparison of DPST with existing MIMO systems. 
Section~\ref{sec:part_Conclusion} draws the conclusions.

\begin{figure*}[t]
\centering
\includegraphics[scale=0.6]{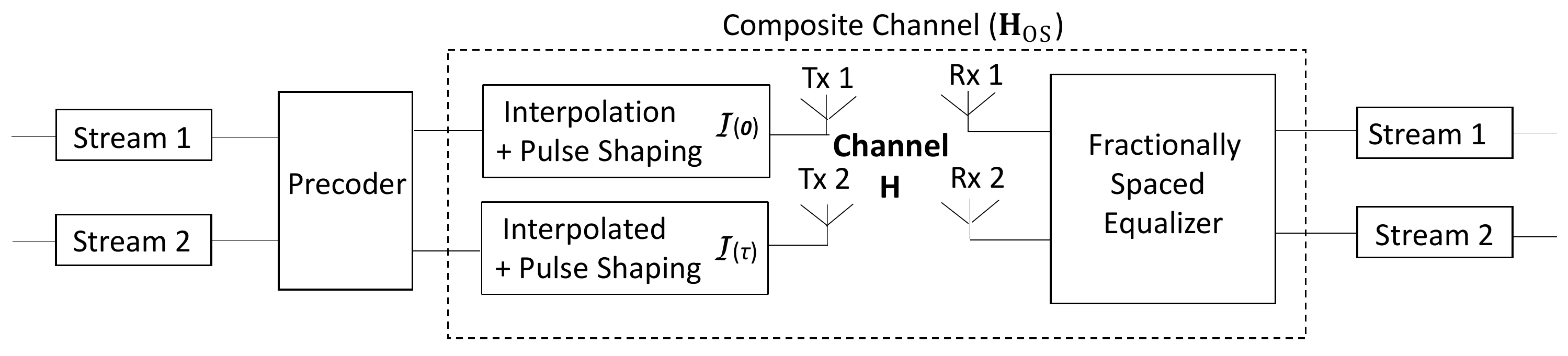}
\caption{DPST block diagram. The fractional delay $\tau$ is introduced to ensure that the effective channel between transmit-receive pairs is well-conditioned.}
\label{fig:DPST_Block}
\vspace{-0.3cm}
\end{figure*}

The following notations are used throughout this paper. 
Bold and lowercase letters denote vectors, 
whereas bold and capital letters denote matrices. 
The notations $()^{-1}$, $()^H$, $()^{*}$ and $()^T$ denote the inverse, Hermitian, conjugate and transpose of a vector or matrix, respectively. The notation $\ast$ also represents the convolution operation. Table \ref{tab:tab_notations} also lists the notions that are used throughput this paper.

\begin{table}[tbph]
\centering
\renewcommand{\arraystretch}{1.4}
\caption{Summary of Notation}
\label{tab:tab_notations}
\scalebox{0.85}{
\centering
\begin{tabular}{|>{\centering\arraybackslash}p{1.7cm}|>{\centering\arraybackslash}p{7cm}|}
\hline \textbf{Notation} & \textbf{Description} \\ \hline
\hline
$T_{s}$ & Symbol period     \\ \hline
$\tau$ & Fractional delay     \\ \hline
$N_{t}$ & Number of transmit antennas     \\ \hline
$N_{r}$ & Number of receive antennas \\ \hline
K & Rician K factor     \\ \hline
$\textbf{s}$ & Transmitted signal matrix     \\ \hline
$\textbf{x}$ & Received signal matrix     \\ \hline
$\textbf{x}_{\rm os}$ & Oversampled signal matrix at receiver     \\ \hline
$\textbf{H}$ & Correlated MIMO channel     \\ \hline
$\textbf{H}_{s}^{\rm LOS}$ & Unit LOS MIMO channel     \\ \hline
$\textbf{H}_{s}^{\rm NLOS}$ & Correlated NLOS MIMO channel     \\ \hline
$\textbf{H}_{\rm os}$ & Composite MIMO channel     \\ \hline
$\textbf{H}_{\rm R}$ & Downsized MIMO channel     \\ \hline
$\textbf{H}_{\rm N}$ & Normalized MIMO channel     \\ \hline
$\textbf{H}_{\rm eq}$ & Precoded MIMO channel     \\ \hline
$\lambda$ & Channel singular value \\ \hline
$h_{i,j}$ & Channel from $j$th transmit antenna to $i$th receive antenna \\ \hline
$a_{i,j}$ & Amplitude gain of $h_{i,j}$ \\ \hline
d & Distance between UE and BS     \\ \hline
$\textbf{R}_{T}$ & Transmit correlation matrix     \\ \hline
$\textbf{R}_{R}$ & Receive correlation matrix     \\ \hline
$\textbf{H}_{w}$ & Spatially white MIMO channel     \\ \hline
$\mathcal{K}$ & MIMO channel condition number     \\ \hline
$M$ & Length of input signal    \\ \hline
$R$ & Transmit oversampling ratio    \\ \hline
$P$ & Receive oversampling ratio    \\ \hline
$\textbf{\ensuremath{\mathcal{I}}}$ & Interpolation matrix     \\ \hline
$\textbf{\ensuremath{\mathcal{I}}}_{\rm Tx}$ & Transmit Interpolation matrix     \\ \hline
$\textbf{\ensuremath{\mathcal{I}}}_{\rm Rx}$ & Receiver Interpolation matrix     \\ \hline
$\textbf{W}$ & Precoding matrix    \\ \hline
$\rho$ & Power scaling ratio     \\ \hline
$\zeta$ & SINR at each receiver branch \\ \hline
$\sigma_{0}^{2}$ & Noise power \\ \hline
$\rm F^{\rm MMSE}$ & MMSE equalizer     \\ \hline
\end{tabular}}
\vspace{-0.2cm}
\end{table}

\section{Channel Correlation Model in Ultra-Dense Small Cell Networks}
\label{sec:part_Correlation}

The performance of a MIMO systems is highly dependent on different channel pairs being spatially uncorrelated. 
In correlated channel scenarios, different transmit-receive antenna pairs will undergo similar channel conditions, 
and as a result the multi-path components corresponding to different pairs may not be resolvable by the UE~\cite{1203167}. 
Thus, correlated channels have reduced degrees of freedom, 
leading to reduced throughputs as detailed in~\cite{1459054}~\cite{892194}.

The Rician multi-path fading model is used in order to capture the impact of LOS communication in ultra-dense small cell networks. We do not consider the channel variations due to UE mobility, 
since ultra-dense small cell networks are designated to provide coverage to static UEs. 

The Rician MIMO channel with $N_{t}$ transmit antennas and $N_{r}$ receive antennas can be represented using an $N_{r} \times N_{t}$ matrix $\textbf{H}$ as
\begin{equation}
  \textbf{H} = \sqrt{\frac{K}{K+1}} {\textbf{H}_{i}}^{\rm LOS} + \sqrt{\frac{1}{K+1}}{\textbf{H}_{s}}^{\rm NLOS},
  \label{eq:LOS_NLOS}
\end{equation}
where $\rm K$ is the Rician K factor, 
and ${\textbf{H}_{i}}^{\rm LOS}$ and ${\textbf{H}_{s}}^{\rm NLOS}$ are the unit LOS MIMO channel matrix and the correlated non-LOS (NLOS) MIMO channel matrix, respectively. 
In~\cite{2015Jafari}, the authors present a distance dependant Rician K factor based on the probability of LOS in micro urban environments, 
which offers a transition from Rician to Rayleigh fading as UEs locate further away from the BSs and the LOS component gradually decades. Worth noting that as the ISD reduces, the LOS component becomes more dominant, which leads to a larger correlation among the different transmit-receive channel pairs. The distance dependant Rician K factor is given as
\begin{equation}
  \rm K=\begin{cases}
      32 & \text{if $d<18 m$} \\
     140.10 \times \exp(-0.107 \times d) & \text{otherwise},
  \end{cases}
  \label{eq:Kfactor}
\end{equation}
where $d$ denotes the distance between the UE and BS. More detailed discussion on derivation of the Rician K factor value is available in ~\cite{2015Jafari}.
To model the correlated NLOS channel matrix, 
the Kronecker model is used, 
which captures the correlation among the channel pairs due to the spacing between antenna elements~\cite{1683382} at both UE and BS. 
We derive $\textbf{R}_{R}$ and $\textbf{R}_{T}$ as the $N_{r} \times N_{t}$ correlation matrices at the receiver and transmitter sides, respectively. 
A detailed discussion on derivation of correlation coefficient between any two channel pairs as a function of the spacing between antenna elements is presented in the Appendix. 

The correlated NLOS MIMO channel ${\textbf{H}_{s}}^{\rm NLOS}$ can therefore be expressed as
\begin{equation}
  {\textbf{H}_{s}}^{\rm NLOS} = {\textbf{R}_{R}}^{\frac{1}{2}} \textbf{H}_{w} {\textbf{R}_{T}}^{\frac{1}{2}},
  \label{eq:3}
\end{equation}
where $\textbf{H}_{w}$ is the $N_{r} \times N_{t}$ spatially white MIMO channel. Plugging (\ref{eq:3}) into (\ref{eq:LOS_NLOS}), the correlated Rician MIMO channel is derived as 
\begin{equation}
  \textbf{H} = \sqrt{\frac{K}{K+1}} {\textbf{H}_{i}}^{\rm LOS} + \sqrt{\frac{1}{K+1}} \textbf{R}_{R}^{\frac{1}{2}} \textbf{H}_{w} \textbf{R}_{T}^{\frac{1}{2}}
  \label{eq:Full_correlated_channel}
\end{equation}

Note that the dependency of spatial correlation on UE-BS distance as well as antenna spacing is attributed to the Rician K factor and the Kronecker product of transmit/receive correlation matrices, respectively. 
Applying singular value decomposition (SVD), 
the $N_{r} \times N_{t}$ channel $\textbf{H}$ can be written as $\textbf{H} = \textbf{U} \mbox{\boldmath{$\Sigma$}} \textbf{V}^{\rm {H}}$ 
where $\textbf{U}$ and $\textbf{V}$ are $N_{r} \times N_{r}$ and $N_{t} \times N_{t}$ unitary matrices, respectively and \mbox{\boldmath{$\Sigma$}} is a $N_{r} \times N_{t}$ diagonal matrix as shown in (\ref{eq:singular}). 
The diagonal entries $\lambda_{l}$ $(l = 1,2,\cdots,L)$ where $L = min(N_{r},N_{t})$ denote the singular values of the channel $\textbf{H}$ in the descending order. 
\begin{equation}
\mbox{\boldmath{$\Sigma$}} = \left(\begin{matrix}
								\lambda_{1} & 0 & \cdots & 0\\
								0 & \lambda_{2} & \cdots & 0\\
								\vdots & \vdots & \ddots & \vdots\\
								0 & 0 & \cdots & \lambda_{L}\\
							 \end{matrix}\right)
\label{eq:singular}
\end{equation}

The condition number ($\mathcal{K}$) of the channel is defined as the ratio of the maximum to minimum singular values $\mathcal{K} = \frac{\lambda_{1}}{\lambda_{L}}$, 
and is referred to as a metric to denote the quality of the independent streams of the wireless channel~\cite{Matrix}. 
$\mathcal{K} \approx 1$ implies no correlation between channel pairs, 
and as long as $\mathcal{K}$ is less than $10$, 
the channel is regarded as well-conditioned, 
and can be leveraged to extract the unitary vectors of the channel for precoding and spatial multiplexing purposes. 

For simplicity - in an $N_{r}=2$ and $N_{t}=2$ setup, 
the singular values corresponding to each transmit antenna can be approximated as
\begin{equation}
\begin{split}
&\lambda_{1} \approx - \frac{|h_{1,1} \ h_{2,2}-h_{1,2} \ h_{2,1}|^{2}}{{a_{1,1}}^{2} \ + \ {a_{1,2}}^{2} \ + \ {a_{2,1}}^{2} \ + \ {a_{2,2}}^{2}} + {a_{1,1}}^{2} \ + \ {a_{1,2}}^{2} \ \\ & + \ {a_{2,1}}^{2} \ + \ {a_{2,2}}^{2} \\ & 
\lambda_{2} \approx \frac{|h_{1,1}h_{2,2}-h_{1,2}h_{2,1}|^{2}}{{a_{1,1}}^{2} \ + \ {a_{1,2}}^{2} \ + \ {a_{2,1}}^{2} \ + \ {a_{2,2}}^{2}}
\end{split}
\end{equation}
where $a_{i,j}$ denotes the amplitude gain of $h_{i,j}$, 
which refers to channel from the $\textit{j}$th transmit antenna to the $\textit{i}$th receive antenna~\cite{5956339}. 
The $2 \times 2$ wireless channel can be decomposed into its $L=2$ singular channels, 
and if both $\lambda_{1}$ and $\lambda_{2}$ are sufficiently large, 
the corresponding parallel channels can be exploited to convey the specific data streams and the capacity $C$ can be computed as the sum of the corresponding singular channels' capacities as follows
\begin{equation}
C = {\rm log_{2}} \ [{\rm {det}} \ (\textbf{I}_{L} + \frac{\zeta}{\sigma_{0}^{2}} \textbf{H}\textbf{H}^{H})]
=\sum\limits_{l=1}^{L} {\rm {log_{2}}} \ (1 + \ \frac{\zeta\lambda_{l}}{L}),
\label{eq:capacity_eq}
\end{equation}
where $\zeta$ is the SINR at each receiver branch and $\sigma_{0}^{2}$ is the receiver noise power~\cite{5956339}.

However, in the correlated $2 \times 2$ MIMO channel deployed in an ultra-dense small cell network, 
$\lambda_{2}$ will be minimum, implying that the number of streams that can be simultaneously transmitted and the achievable rates are limited. This elaborates the performance degradation of MIMO spatial multiplexing in ultra-dense small cell networks in comparison to non ultra-dense small cell networks. In next section, we propose a new transmission technique to tackle this issue.

\section{Delayed Pulse Shaping Transmission}
\label{sec:part_DPST}

In the $2\times2$ MIMO system, 
DPST works by shaping the data stream of the second transmit antenna with respect to first transmit antenna such that in downlink communications the multi-path components corresponding to the second transmit antenna arrive at the UE at a later time instant than those corresponding to the first transmit antenna, 
aiming to enhance the diversity among the \textit{fractionally delayed} multi-path components of the closely placed transmit antennas. Fig.~\ref{fig:DPST_Block} shows the block diagram of DPST in a $2\times2$ setup.

The impact of the \textit{deterministic fractional delay} on enhancing the diversity of the channel by means of ISI generation is similar to the outcome of faster than Nyquist (FTN) signalling~\cite{6479673}~\cite{4777625}~\cite{1231648}. 
In FTN communications, a non-orthogonal sampling kernel is used to allow for signalling above the Nyquist limit, 
and hence data streams are transmitted at a rate above the Nyquist symbol period~\cite{4777625}~\cite{1231648}, eventually improving the communication rates at the cost of a more complicated receiver design to combat the introduced ISI through oversampling. In the traditional cases, by using an orthogonal sampling kernel, 
it means that there is only a single non-zero component of the transmitted signal. However, in FTN this property is traded for increased rates.

\begin{figure}[t]
\centering
\includegraphics[scale=0.66]{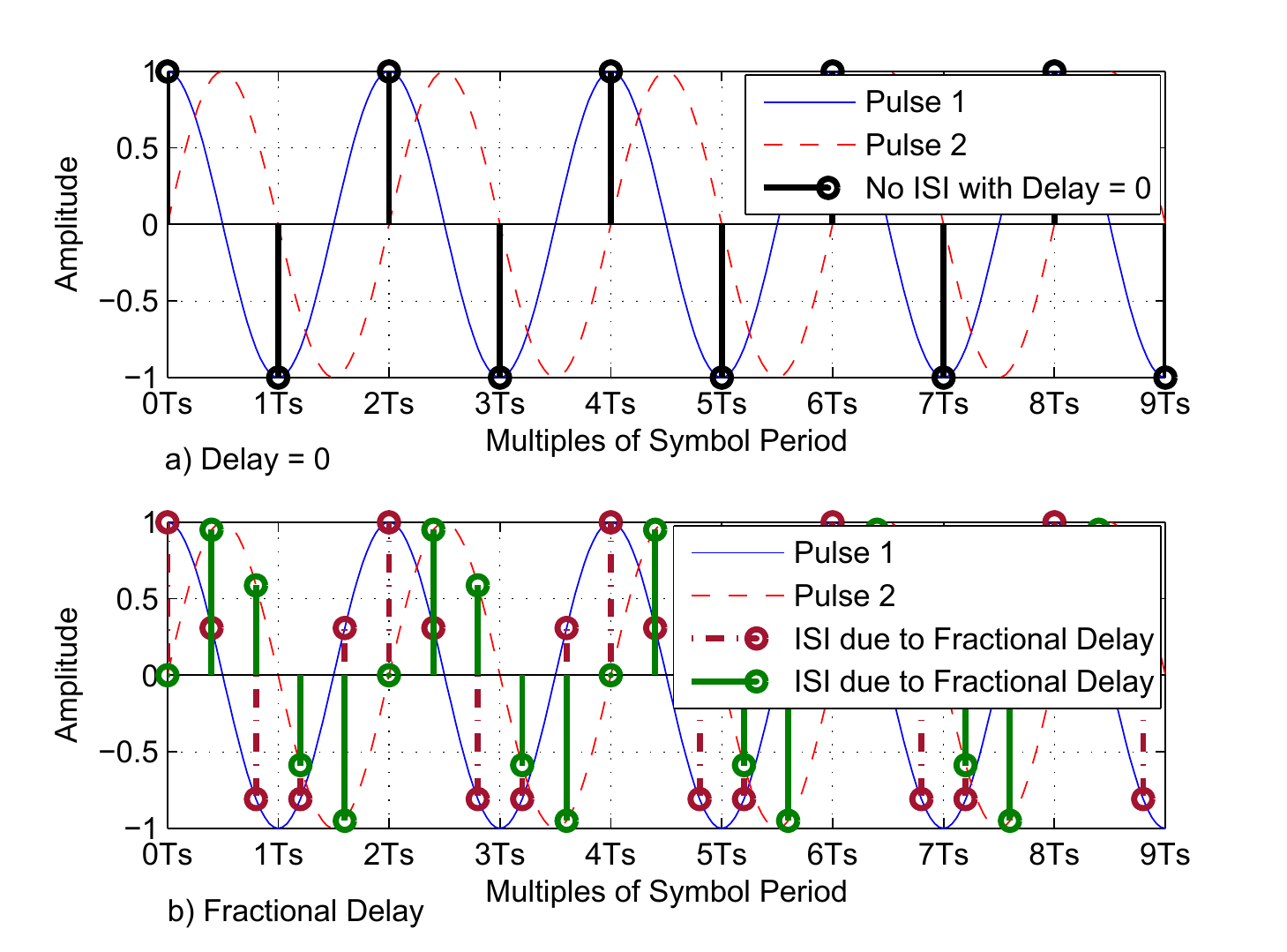}
\caption{Integral and non-integral sampling effect on DPST.}
\label{fig:FTN_DPST}
\vspace{-0.3cm}
\end{figure}

The additional \textit{fractional delay} that is applied to the transmission of the second antenna in DPST also generates \textit{deterministic ISI} in the system~\cite{6620766}, 
which suggests the analogy between DPST and FTN in terms of generation of controlled ISI. 
Fig.~\ref{fig:FTN_DPST} intuitively shows the implication of DPST. 
In Fig.~\ref{fig:FTN_DPST}a, 
the signals are sampled at integral multiples of symbol period with an orthogonal sampling kernel, 
while in Fig.~\ref{fig:FTN_DPST}b, 
signals are sampled at non-integral multiples of symbol period. 
In the latter case and in contrast with the former, 
at each sampling instant, there are multiple non-zero samples viewed by the sampling kernel, 
which shows how at each sampling point the pulses interfere. 
As a result of the generated ISI in the system, 
the sampled channel impulse response $h[nT_{s}]$ will no longer follow the Nyquist zero ISI criterion presented in (\ref{eq:ISI_Nyq}) with $n$ being an integer. 
The generated ISI is then exploited to increase the diversity of the wireless channel seen between the transmit and receive antennas pairs, 
implying reduced correlation among different channel pairs as viewed by the receiver.
\begin{equation}
h[nT_{s}] =
\begin{cases}
1 \ \ \ n = 0 \\
0 \ \ \ n \neq 0
\end{cases}
\label{eq:ISI_Nyq}
\end{equation}

\subsection{MIMO Link Model with Pulse Shaping}
\label{ssec:Transmitter}

The deterministic transmission delay $\tau$ must be a fraction of the transmitted signal period $T_s$, 
i.e., $0 < \tau < T_s$, where $T_s$ is normalised to be~$1$. 

The received signal can be expressed as
\[
\bea{ccc}
\left[\bea{c}
x_{1}(t) \\ x_{2}(t)
\ena\right]
& = &
\left[\bea{cc}
h_{1,1}(t) & h_{1,2}(t) 
\\ h_{2,1}(t) & h_{2,2}(t)
\ena\right]
\ast
\left[\bea{c}
s_{1}(t) \\ s_{2}(t+\tau)
\ena\right]
\vspace{0.2cm}
\\
\bx(t) & = & \bH(t) \ast \bs(t)
\ena
\]
where $\textbf{H}(t)$ is the continuous time version of the $2\times2$ correlated MIMO channel, 
$s_{1}(t)$ and $s_{2}(t)$ are the signals transmitted by first and second transmit antennas, respectively, 
while $x_{1}(t)$ and $x_{2}(t)$ are the signals received by first and second receive antennas, respectively.
In order to implement the fractional delay $\tau$, 
the transmit data streams are oversampled/interpolated as shown in the following
\beqa
\left[\bea{@{}c@{}}
x_{1}(t) 
\\ x_{2}(t) 
\ena\right]
\hspace{-0.05cm}
=
\hspace{-0.05cm}
\left[ \bea{@{}cc@{}}
h_{1,1}(t) & h_{1,2}(t) \\ 
h_{2,1}(t) & h_{2,2}(t) 
\ena\right] 
\hspace{-0.05cm}
\ast
\hspace{-0.05cm}
\left(
\left[\bea{@{}cc@{}}
\cI(0)s_{1}(t) & \hspace{-0.3cm} 0 
\\ 0 & \hspace{-0.3cm} \cI(\tau)s_{2}(t)
\ena\right]
\right)
\label{eq:21}
\enqa
where $\textbf{\ensuremath{\mathcal{I}}}(0)$ and $\textbf{\ensuremath{\mathcal{I}}}(\tau)$ are $N \times M$ interpolation matrices that are applied to the first and second transmit antennas, respectively. 
$\cI_{\rm Tx}$ refers to the $2N \times 2M$ interpolation matrix encompassing the interpolation matrices at both transmit antennas where $M$ denotes the length of input signal with period $T_{s}$, $R$ is the oversampling ratio at the transmitter and $N = M R$~\cite{5706377}.

The modelling of oversampled analogue signals follows~\cite{5706377}, and the elements of $N\times M$ interpolation matrix are obtained as
\begin{equation}
\ensuremath{\mathcal{I}}_{nm} = {\rm {sinc}} \ \left(\frac{n(\frac{T_{s}}{N})+ \tau -m(\frac{T_{s}}{M})}{\frac{T_{s}}{M}}\right) \quad
\begin{cases}
\textit{m} = 1,2,...,M \\ \textit{n} = 1,2,...,N 
\end{cases}
\label{eq:22}
\end{equation}

It is evident that the interpolation matrix corresponding to transmit antenna~1 has no time offset, 
while the one corresponding to transmit antenna~2 is offset by time $\tau$ to account for delayed pulse shaping. 

The delay $\tau$ plays a key role in DPST performance. 
When the delay $\tau=0$, 
the expression presented in (\ref{eq:21}) collapses into the initial ill-conditioned wireless channel $\bH$. 
Alternatively, for integral multiples of $\tau=k  T_s$ ($k = 1,2,...$), 
integral DPST provides a cyclic shift of transmitted data streams at the receiver.  
For fractionally delayed values of $\tau \neq k  T_s$ ($k = 1,2,...$), 
the matrix $\ensuremath{\mathcal{I}}(\tau)$ becomes a block diagonal matrix operating on the streams of the input sequence $\{s_2[1], \, s_2[2], \, \cdots s_2[k]\}$ which correspond to the second transmit antenna. 
As a result, the signals corresponding to adjacent antenna elements are shaped with distinct pulse shaping filters. Note that this will introduce ISI in time domain, 
which increases the order of the composite channel $\textbf{H}_{rm os}$ between the transmit and receive streams, 
as shown in Fig.~\ref{fig:DPST_Block}. 
Also note that the fractional delay specifies that the columns of the interpolation matrix $(\ensuremath{\mathcal{I}}\tau)$ are not orthogonal any more, which leads to different non-zero pulses at sampling points. Therefore, linear combinations of delayed $s_{2}(t)$ will lead to ISI. It is worth noting that the transmitted signals by transmit antenna~1 and transmit antenna~2 are not matched any more, and thus the corresponding signal degradation is taken into account. Moreover, in case of sinc pulse shaping, even a very small fractional delay introduces ISI terms, 
which is exploited to enhance the diversity among different channel pairs.

\subsection{Receiver Design Considerations - Fractionally Spaced Equalisation}
\label{ssec:Receiver}

The sinc interpolation shown in ~(\ref{eq:21}) can be interchanged and alternatively represented as
\beqa
\left[\bea{@{}c@{}}
x_{1}(t) 
\\ x_{2}(t) 
\ena\right]
=
\left( 
\left[ \bea{@{}cc@{}}
h_{1,1}(t) & h_{1,2}(t) \\ 
h_{2,1}(t) & h_{2,2}(t) 
\ena\right] 
\ast
\left[\bea{@{}c@{}c@{}}
\cI(0) & 0 
\\ 0 & \cI(\tau)
\ena\right]
\right)
\left[\bea{@{}c@{}}
s_{1}(t) \\ 
s_{2}(t) 
\ena \right]
\nonumber
\enqa

Assuming that ultra-dense small cell networks are interference-limited, 
the noise term is ignored and so the received signal at receive antennas 1 and 2 can be rewritten as
\[
\left[\bea{c}
x_{1}(t) \\ 
x_{2}(t) 
\ena \right] = 
\left[ \bea{cc}
\cI(0) \ast h_{1,1}(t) & \cI(\tau) \ast h_{1,2}(t) \\
\cI(0) \ast h_{2,1}(t) & \cI(\tau) \ast h_{2,2}(t)
\ena\right] 
\left[\bea{c}
s_{1}(t) \\ 
s_{2}(t) 
\ena \right]
\]

At the receiver side, 
to comply with DPST requirement, 
the receiver must operate at a rate significantly greater than the symbol period. 
This is in order to be consistent with the interpolation stage at the transmitter as well as to mitigate the ISI. 
For this purpose, fractionally spaced equaliser (FSE) is used, 
which is a finite impulse response (FIR) filter and the spacing between its taps is a fraction of symbol period. 
FSE is implemented by sampling the received signals $P$ times (typically $P\geq 2$) within the time interval $t \in \left(kT_s, \, [k+1]T_s\right]$ 
where $t = (k+\frac{1}{P})T_s$ and is stacked as a $P\times1$ vector, i.e.,
\[
\bea{ccc}
\left[ \bea{@{}c@{}}
x_{1}(t=kT_s) \\ \vdots \\ x_{1}(t=(k+\frac{P-1}{P})T_s)
\ena\right]
& = &
\underbrace{
\cI_{\rm R}
\left[\bea{@{}cc@{}}
\cI(0) \ast h_{1,1}(t) \\ \cI(\tau) \ast h_{1,2}(t)
\ena\right]^{T}}_{\bH_{1,os}[k]}
\left[\bea{@{}cc@{}}
s_{1}(t) \\ 
s_{2}(t) 
\ena \right]
\vspace{0.3cm}
\\
\bx_{1,os}[k] & = & \bH_{1,os}[k] \ \bs[k]
\ena
\]

\[
\bea{ccc}
\left[ \bea{@{}c@{}}
x_{2}(t=kT_s) \\ \vdots \\ x_{2}(t=(k+\frac{P-1}{P})T_s)
\ena\right]
& = &
\underbrace{
\cI_{\rm R}
\left[\bea{@{}cc@{}}
\cI(0) \ast h_{2,1}(t) \\ \cI(\tau) \ast h_{2,2}(t)
\ena\right]^{T}}_{\bH_{2,os}[k]}
\left[\bea{@{}cc@{}}
s_{1}(t) \\ 
s_{2}(t) 
\ena \right]
\vspace{0.3cm}
\\
\bx_{2,os}[k] & = & \bH_{2,os}[k] \ \bs[k]
\ena
\]
where $\textbf{s}[k] = \textbf{s}(t = k T_{s})$ and $\textbf{x}_{1,os}$ and $\textbf{x}_{2,os}$ are the interpolated signals at the first and second receive antennas, respectively, 
and represent $P \times 1$ vectors. $\cI_{R}$ is also the $(N \times P) \times N$ oversampling matrix at each receive antenna. 

\begin{equation}
\ensuremath{\mathcal{I}}_{pn} = {\rm {sinc}} \ \left(\frac{p(\frac{T_{s}}{P \times N}) - n(\frac{T_{s}}{N})}{\frac{T_{s}}{N}}\right) \quad
\begin{cases}
\textit{n} = 1,2,...,N \\ \textit{p} = 1,2,...,P \times N 
\end{cases}
\label{eq:sinc_rx}
\end{equation}

Note that for consistency with the interpolated pulse shaping at transmitter, 
\textit{sinc interpolation} as shown in (\ref{eq:sinc_rx}) is also exploited at the receiver. 
From a receiver perspective, 
the signals from the two transmit antennas are observed at $kT_{s}$ and $kT_{s}+\tau$ ($k = 1,2,...$), respectively. 
It is important to realise that the triggered diversity can be only extracted by the receiver if it is operating in an oversampled domain. Moreover, exploiting an oversampled receiver allows the UE to suppress the ISI using an equalizer. 

Stacking the $P \times 1$ vectors $\textbf{x}_{1,os}[k]$ and $\textbf{x}_{2,os}[k]$ from both receive antennas, 
we obtain
\beq
\bx_{os}[k] =
\left[ \bea{c}
\bx_{1,os}[k]\ \\ \bx_{2,os}[k]
\ena\right]
=
\left[ \bea{c}
\bH_{1,os} \\ \bH_{2,os}
\ena\right]
\ast
\bs[k]
= 
\bH_{os} \ast \bs[k]
\label{eq:27}
\enq

It can be understood from~(\ref{eq:27}) that due to fractionally delayed interpolated pulse shaping, 
the diversity of the channel is enhanced because the columns of $\textbf{H}_{1,\rm {os}}$ and $\textbf{H}_{2,\rm {os}}$ are independent of each other. 
This indicates that the rank of the composite channel $\textbf{H}_{\rm os}$ is greater than 1.
It is also noticed that the composite channel matrix $\textbf{H}_{\rm os}$ is a tall matrix, 
which becomes a full rank column matrix subject to the sufficient pulse shaping diversity.
Moreover, it has to be indicated that while SINR at a given sampling instant can be degraded due to pulses interfering with each other, 
the overall SINR after receiver equalisation can still be large enough by exploiting the pulse diversity to allow decoding. 
 
From the receiver point of view, 
the combined effect of DPST and FSE enhances the degrees of freedom and reduces the correlation between different pairs of the composite channel $\textbf{H}_{\rm os}$. 
This composite channel $\textbf{H}_{os}$ can be represented as
\[
\textbf{H}_{\rm os} = \textbf{\ensuremath{\mathcal{I}}}_{\rm Rx} (\textbf{H} \ast \textbf{\ensuremath{\mathcal{I}}}_{\rm Tx}),
\label{eq:28}
\]
where $\textbf{\ensuremath{\mathcal{I}}}_{\rm Tx}$ and $\textbf{\ensuremath{\mathcal{I}}}_{\rm Rx}$ are $2N \times 2M$ and $2P \times 2N$ matrices that are comprised of the interpolation matrices at both antennas at the transmitter and receiver sides, respectively. 
It is worth reminding that the deterministic fractional delay is incorporated in $\textbf{\ensuremath{\mathcal{I}}}_{\rm Tx}$.

\subsection{Matrix Dimension Reduction and Power Normalization}

Due to the interpolation stages involved at both transmitter and receiver sides, 
the composite channel matrix $\textbf{H}_{\rm os}$ is of larger dimensions than $\textbf{H}$. 
To downsize $\textbf{H}_{\rm os}$ with respect to $\textbf{H}$, 
the composite channel is decomposed using SVD as $\textbf{H}_{\rm os} = \textbf{U}_{\rm os} $\mbox{\boldmath{$\Sigma$}}$_{\rm os} \textbf{V}_{\rm os}^{\rm {H}}$, 
and since $\textbf{H}$ refers to a $2\times2$ MIMO channel matrix, 
$\textbf{U}_{\rm os}$ and $\textbf{V}_{\rm os}$ only take the first two columns and rows, respectively, in the sequel. 
\[
\textbf{H}_{\rm R} = \textbf{U}_{\rm os}(: \ ,1:2)^{\rm {H}} \ \textbf{H}_{\rm os} \ \textbf{V}_{\rm os}(1:2 \ , \ :)^{\rm {H}}
\]
The downsized channel $\textbf{H}_{\rm R}$ must also be normalised with respect to the initial channel $\textbf{H}$. The normalised channel is then obtained as
\[
\textbf{H}_{\rm N} = \textbf{H}_{\rm R} \times \frac{\parallel \textbf{H} \parallel_{2}}{\parallel \textbf{H}_{\rm R} \parallel_{2}},
\]
where $\parallel$ $\parallel$ is the Frobenius norm.

Prior to downsizing, 
there are still some redundancy in the channel matrix $\textbf{H}_{\rm os}$. 
However, when the dimensions of the channel is reduced, 
the residual small singular values are removed, 
and, therefore, the condition number of the channel $\textbf{H}_{\rm R}$ witnessed by the receiver is improved. 
Overall, the diversity enhancement of the virtual MIMO channel $\textbf{H}_{\rm N}$ is attributed to 
\begin{itemize}
\item 
the introduction of deterministic fractional delay realised though distinct interpolated pulse shapes,
\item 
the higher order channel observed at the receiver due to fractionally spaced equaliser and the elimination of the residual singular values.
\end{itemize}

\subsection{Precoding and Detection}
\label{ssec:prec_detec}

As discussed in \ref{ssec:Transmitter} and \ref{ssec:Receiver}, 
the virtual channel $\textbf{H}_{\rm N}$ benefits from enhanced diversity. 

At the transmitter, 
in order to perform precoding, 
we do SVD as $\textbf{H}_{\rm N} = \textbf{U}_{\rm N} $\mbox{\boldmath{$\Sigma$}}$_{\rm N} \textbf{V}_{\rm N}^{\rm {H}}$, 
and extract the first two columns of $\textbf{V}_{\rm N}$ to generate the precoding matrix denoted by $\textbf{W}$ 
where $\textbf{W} = \textbf{V}_{\rm N}(: \ ,1:2)$. 
The precoding matrix is then subject to power scaling since it must not violate the BS transmission power constraint, 
i.e.,
\[
\textbf{W} = \sqrt{P_{ \rm BS}} \times \rho \times \textbf{W},
\]
where $P_{\rm BS}$ is the BS transmit power and $\rho$ is the power scaling ratio. 
The precoded channel $\textbf{H}_{\rm eq}$ is then defined as $\textbf{H}_{\rm eq} = \textbf{H}_{\rm N} \textbf{W}$. 

At the receiver side, 
minimum mean squared error (MMSE) equaliser is exploited, 
which is defined as
\beq
\bF^{\rm {MMSE}}
=
\left[ \bea{cc} f_{1,1} & f_{1,2} \\ f_{2,1} & f_{2,2} \ena\right]  = \textbf{H}_{\rm eq}^{\rm {H}}(\textbf{H}_{\rm eq}\textbf{H}_{\rm eq}^{\rm {H}} + \textbf{$\Phi$} + \textit{N}_{0}\textbf{I})^{-1},
\enq
where $\textit{N}_{0}$ is the noise power 
and $\Phi$ is the inter-cell interference covariance matrix at the receiver denoted as $\textbf{$\Phi$} = \rm {E}\{\textbf{v} \textbf{v}^{H}\}$ 
where $\textbf{v} = [v_{1} \ v_{2}]^{T}$. 

The SINR corresponding to data streams 1 and 2 are then computed as
\begin{equation*}
SINR_{x_{1}}^{\rm {Receive}} = \frac{|f_{1,1}h_{1,1} + f_{1,2}h_{2,1}|^{2}}{I_{x_{1}} + N_{0}(|f_{1,1}|^{2} + |f_{2,1}|^{2})},
\end{equation*}
\begin{equation*}
SINR_{x_{2}}^{\rm {Receive}} = \frac{|f_{2,1}h_{1,2} + f_{2,2}h_{2,2}|^{2}}{I_{x_{2}} + N_{0}(|f_{1,2}|^{2} + |f_{2,2}|^{2})}
\end{equation*} 
where ${\rm {I}}_{x_{1}}$ and ${\rm {I}}_{x_{2}}$ refer to the interferences at first and second receive antennas, 
and are defined as $|f_{1,1}h_{1,2} + f_{1,2}h_{2,2}|^{2}\ + \ {\rm {E}}\{|f_{1,1}v_{1}+f_{1,2}v_{2}|^{2}\}$ and $|f_{2,1}h_{1,1} + f_{2,2}h_{2,1}|^{2}\ + \ {\rm {E}}\{|f_{2,1}v_{1}+f_{2,2}v_{2}|^{2}\}$, respectively.

\vspace{0.2cm}
 
The transmitted data streams can be thus estimated using $\textbf{F}^{\rm {MMSE}}$ operating on the received signals as
\begin{equation}
\hat{\bs}[k] = \bF^{\textrm{MMSE}} (\bU_{\textrm{os}}^{\rm H} \bH_{\textrm{os}} \ast \bs[k])
\label{eq:received_sig}
\end{equation}

Despite that ultra-dense small cell networks are assumed to be interference limited, 
the DPST requirement for a receiver to operate in an oversampled domain according to~\ref{ssec:Receiver}
can give rise to the question of the oversampling impact on the noise signal and accordingly on the overall performance. 
Note that we have a band limited signal that is corrupted by thermal and quantisation noise. 
However, oversampling only modifies the quantisation noise, 
and does not alter any thermal noise terms and hence the noise spectral density does not change.

\subsection{Differentiation with Cyclic Delay Diversity}
\label{ssec:part_CDD}

Cyclic delay diversity (CDD)~\cite{4656965}~\cite{989781}~\cite{4109390} is known as a diversity technique used in long term evolution (LTE) for spatial multiplexing applications aiming to enhance the diversity between different data streams. In CDD, different antennas transmit a cyclically shifted version of the $\textit{same signal}$ to achieve a diversity gain~\cite{5426062}. The cyclic shift in time domain is equivalent to phase shift in frequency domain, 
and offers the same impact as frequency diversity.

It is important to emphasise that DPST differs from CDD. On the contrary to CDD, in DPST each antenna transmits its own individual signal, which is modulated with a distinct pulse shape and then transmitted with a fractional real delay with respect to its former antenna, requiring to redesign the precoder and the receiver. According to \cite{4212739}, in CDD, the signal is not truly delayed, but cyclically shifted among the transmit antennas. While it might be projected that the fractional delay in time domain is translated into phase shift in frequency domain, 
the functionality of DPTS is different from techniques that aim to mitigate the correlation through phase rotation precoding techniques~\cite{5768230}. Owing to the fractional delay and interpolated pulse shaping, 
DPST injects deterministic ISI in order to increase the diversity among different channel pairs followed by an oversampled receiver to extract such diversity. However, if the delay is an integral of symbol period $T_{s}$, DPST can be interpreted as a variation of cyclic delay diversity (CDD), which does not lead to spatial multiplexing gain. The BS pulse shaping filter shapes the pulses so that the zero-crossings at the output of the receiving filter take place at integral multiples of $T_{s}$. Therefore, when sampling is performed at integral multiples of $T_{s}$, only one pulse is non-zero and the rest are zeros, complying with the original rule of orthogonality. 

It is also necessary to point out that while exploiting antenna polarisation at the transmitter is an effective technique in MIMO systems, 
it is usually limited to two transmit antennas~\cite{1192168}. 
Conversely, DPST can be applied to larger MIMO systems once the set of optimised fractional delays are acquired and applied to the corresponding transmit antennas.

\section{Optimisation of Deterministic Delay in DPST}
\label{sec:part_Optimization}

To optimise the deterministic delay, 
we suggest that the performed FSE at the receiver is not considered as part of DPST delay optimisation. 
FSE is integrated into the DPST receiver design, 
and, therefore, is treated independently. 

The optimisation aims to diagonalise the covariance matrix $\textbf{R}_{\rm x}$ corresponding to the fractionally delayed interpolated pulse shaped channels associated with both transmit antennas $\textbf{H}_{\rm {Tx}}$, 
i.e., a fully diagonal covariance matrix has only non-zero elements on its main diagonal and the rest of its elements are zero. 
Thinking in this line, 
the optimum delay aims to maximise each of the diagonal elements of the $\textbf{R}_{\rm {x}}$ to $1$, 
while minimising all non-diagonal elements. 
As a matter of fact, the optimisation attempts to jointly maximise the auto-correlation, 
while minimising the cross-correlation of the transmit antennas fractionally delayed interpolated pulse shaped channel covariance matrix.
\beqa
\bH_{\rm Tx,1} & = & \bH(:\ ,\ 1) \ \ast \ \ensuremath{\mathcal{I}}(0)
\nonumber \\
&=&
\left[ \bea{@{}c@{}} h_{1,1} \\ h_{2,1} \ena\right]^{\rm T} 
\ast 
\left[ \bea{cccc} \ensuremath{\mathcal{I}}_{11}(0) & \ensuremath{\mathcal{I}}_{12}(0) & \cdots & \ensuremath{\mathcal{I}}_{1M}(0) \\ \ensuremath{\mathcal{I}}_{21}(0) & \ensuremath{\mathcal{I}}_{22}(0) & \cdots & \ensuremath{\mathcal{I}}_{2M}(0) \\
\vdots & \vdots & \ddots & \vdots \\
\ensuremath{\mathcal{I}}_{N1}(0) & \ensuremath{\mathcal{I}}_{N2}(0) & \cdots & \ensuremath{\mathcal{I}}_{NM}(0)
\nonumber
\ena\right]
\enqa

\beqa
\bH_{\rm Tx,2} & = & \bH(:\ ,\ 2) \ \ast \ \ensuremath{\mathcal{I}}(\tau)
\nonumber \\
&=&
\left[ \bea{@{}c@{}} h_{1,2} \\ h_{2,2} \ena\right] ^{\rm T}
\ast 
\left[ \bea{cccc} \ensuremath{\mathcal{I}}_{11}(\tau) & \ensuremath{\mathcal{I}}_{12}(\tau) & \cdots & \ensuremath{\mathcal{I}}_{1M}(\tau) \\ \ensuremath{\mathcal{I}}_{21}(\tau) & \ensuremath{\mathcal{I}}_{22}(\tau) & \cdots & \ensuremath{\mathcal{I}}_{2M}(\tau) \\
\vdots & \vdots & \ddots & \vdots \\
\ensuremath{\mathcal{I}}_{N1}(\tau) & \ensuremath{\mathcal{I}}_{N2}(\tau) & \cdots & \ensuremath{\mathcal{I}}_{NM}(\tau)
\nonumber
\ena\right]
\enqa
where the components of $\ensuremath{\mathcal{I}}_{nm}(\tau)$ are computed based on (\ref{eq:22}). 
The corresponding covariance matrix is thus defined as 
\beqa
\bH_{\rm {Tx}}
& = &
\left[ \bea{cc} \bH_{\rm Tx,1} & \bH_{\rm Tx,2} \ena\right] ^{T}
\label{eq:full_interpolation}
\enqa
\beqa
\bR_{\rm x}(\tau) = \bH_{\rm Tx} \  {\bH_{\rm Tx}}^{\rm T}
\enqa
\beqa
=\left[ \bea{cccc} \textbf{r}_{x_{1,1}}(\tau) & \textbf{r}_{x_{1,2}}(\tau) & \cdots & \textbf{r}_{x_{1,P}}(\tau) \\ \textbf{r}_{x_{2,1}}(\tau) & \textbf{r}_{x_{2,2}}(\tau) & \cdots & \textbf{r}_{x_{2,P}}(\tau) \\
\vdots & \vdots & \ddots & \vdots \\
\textbf{r}_{x_{P,1}}(\tau) & \textbf{r}_{x_{P,2}}(\tau) & \cdots & \textbf{r}_{x_{P,P}}(\tau)
\nonumber
\ena\right]
\enqa

The optimisation problem can be formulated as
\begin{equation*}
{\rm {arg max}} \ \ \textbf{r}_{x,i,i} = 1 \ \forall \ i = 1,2,...,P
\end{equation*}
\hspace{4cm}
\begin{equation}
\textit{subject to} \ 
\begin{cases}
{\ \ \rm {arg min}} \ \ \textbf{r}_{ x,i,j} \ \ \forall \ i,j = 1,2,...,P  \ \ i\neq j \\
\ \ \tau < T_{s}
\end{cases}
\end{equation}

For antenna arrays of more than two antennas, 
the optimisation is modified since adjacent transmit antennas could be subject to non-identical fractional delays. 
Consequently, the optimisation becomes a multi-variate one that aims at finding the Pareto set of fractional delays corresponding to each transmit antenna that leads to the channel with least correlation. 
In this case, (\ref{eq:full_interpolation}) can be obtained as 
\beqa
\bH_{{\rm Tx}, u \ (u \ = \ 2,3,...,U)} = \bH(:\ ,\ u) \ \ast \ \ensuremath{\mathcal{I}}(\tau_{u})
\nonumber \\
\bH_{\rm {Tx}}
=
\left[ \bea{cccc} \bH_{\rm Tx,1} & \bH_{\rm Tx,2} & \cdots & \bH_{{\rm Tx},K} \ena\right] ^{T}
\nonumber
\enqa
where $\tau_{u}$ refers to the fractional delay imposed to the transmission of $u^{th}$ transmit antenna 
and $U$ is the total number of transmit antennas.

Note that an optimum delay will result in a channel, 
whose singular values all exist within the interval $[1 - \epsilon \ , \ 1 + \epsilon ]$ where $\epsilon \in (0,1)$. 
This is a hard problem since it involves SVD as one of the steps and we cannot find a polynomial to formulate the problem. Moreover, since an existing NP-complete problem cannot be encoded into this, 
we can conclude that this is an NP hard problem. 
Considering that this is an NP hard problem, 
there is no efficient, general purpose methods \cite{7270991} \cite{7496967} that can be exploited. 
Since the fractional delay $\tau$ is a one dimensional element and is bounded by cyclic prefix duration, 
an exhaustive search over discrete values of $\tau$ can provide us with a reasonably accurate estimate.

\section{Simulation Results}
\label{sec:part_Simulation}

We consider a single tier hexagonal layout consisting of seven small cell BSs in a $500m \times 500m$ scenario with different ISDs to observe the impact of DPST on various degrees of network densification.  
The central cell is designated as the serving cell and the remaining six cells are considered as interferers. 
We only consider downlink communication and assume that all the small cell BSs operate in the 2 GHz band. 
Macrocell BSs are assumed to operate in a different frequency band and are not considered in this performance evaluation. 
To capture the enhanced spatial correlation due to densification, 
we take into account the ISDs of 20~m, 50~m and 100~m. 
Each small cell consists of an array of two and four transmit antennas, 
and only serves a single UE in one frequency resource. 
The UE also has two and four antennas, thus forming $2\times2$ and $4\times4$ MIMO systems, respectively. 
We assume a fixed spacing of half wavelength between antenna elements of both arrays at UE and BS.
The input signal consists of 10 samples with an oversampling ratio of 2. 
We concentrate on a single frequency resource case,
and antenna gain, path loss, lognormal shadowing and multi-path Rician fast fading are included in SINR computation. 
The path loss model that is used is the microcell urban model defined in~\cite{3gpp}, 
which includes both the LOS and NLOS components. 
At the receiver, MMSE receive filter is used. 
Note that closed-loop precoding is considered.

In the following, we first show how spatial correlation as a result of densification degrades the effective SINR and UE throughput. 
We then provide the simulation results for DPST, 
and show how it can considerably enhance the effective SINR as well the UE throughput.

\subsection{Performance Degradation versus Densification}

To show the performance degradation due to densification, 
we compare the effective SINR and UE throughput CDFs of $2\times2$ and $4\times4$ MIMO systems under both uncorrelated and correlated multi-path fading channel conditions. 
When considering uncorrelated multi-path channel, 
the channel taps are Rayleigh distributed, 
while for correlated scenarios, 
the channel model presented in Section~\ref{sec:part_Correlation} is used, 
which captures the impacts of UE-BS distance and the spacing between antenna elements on spatial correlation.

\begin{figure}
	\centering
	\subfigure[Effective SINR CDF.]{\includegraphics[scale=0.595]{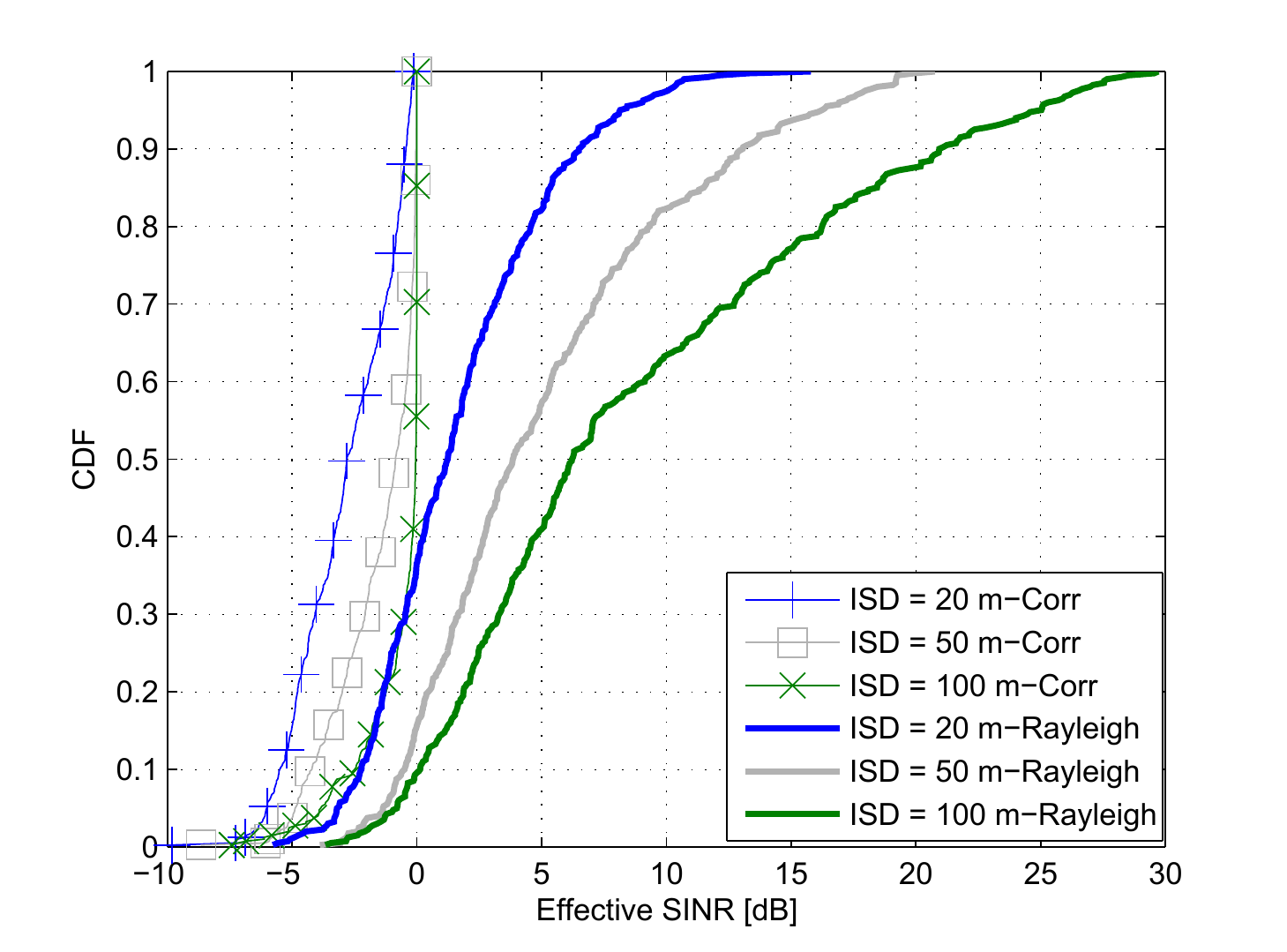}
	\label{fig:SINR_1}}
	\\
	\subfigure[UE throughput CDF.]{\includegraphics[scale=0.595]{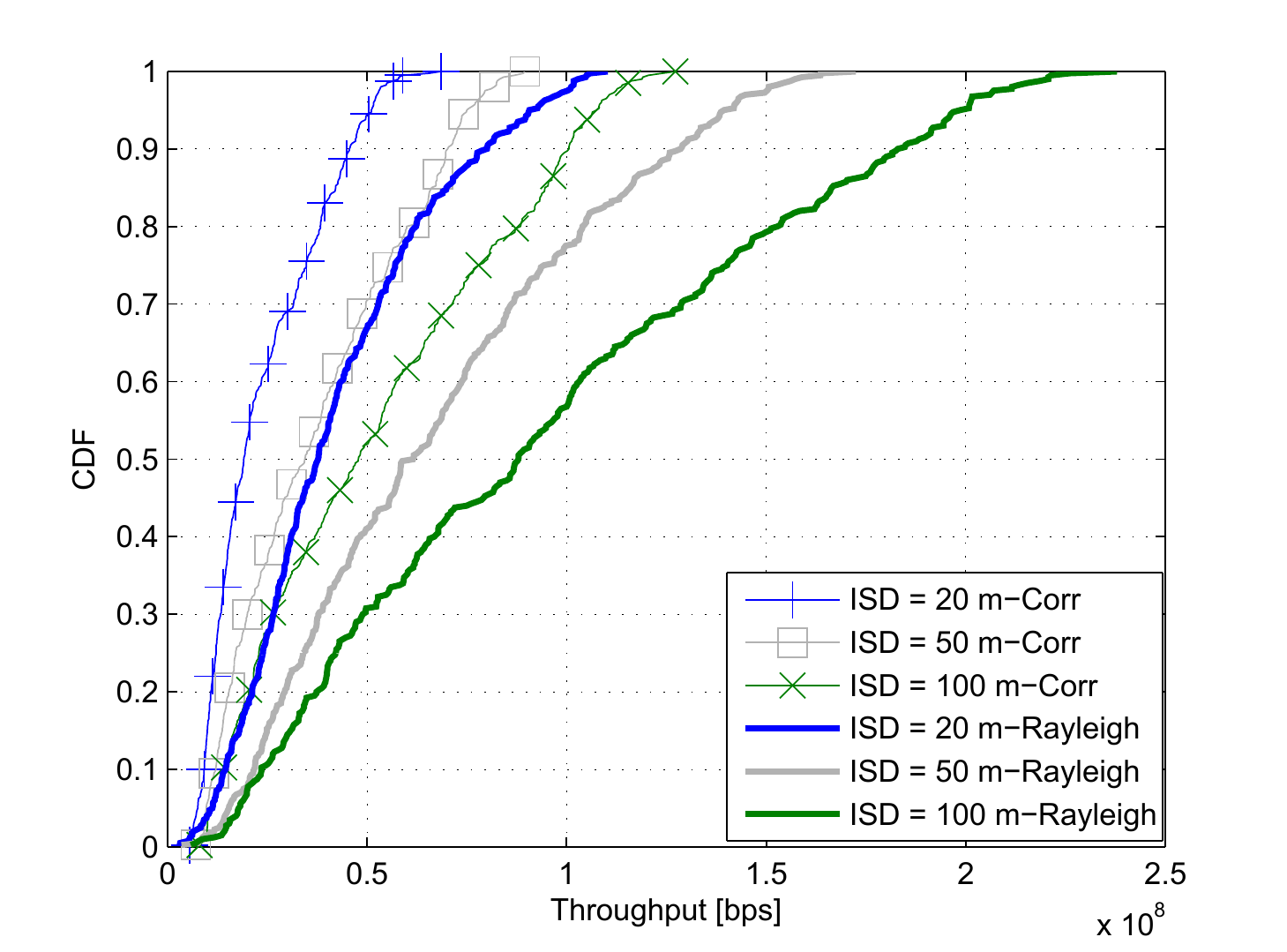}
	\label{fig:Throughput_1}}
	\caption{SINR and throughput CDFs comparison of correlated and Rayleigh channels for a 2x2 MIMO.}
	\label{fig:2x2MIMO_1}
\end{figure}

Fig.~\ref{fig:SINR_1} compares the effective SINR CDF of a $2\times2$ MIMO system 
when using uncorrelated and correlated multi-path fading channels at different ISDs. 
When considering uncorrelated scenarios,
it can be observed that the SINR CDF worsens with the lower ISD. 
This is because the path loss of interfering signals transit from NLOS to LOS interference with the densification, and  therefore, the interference grows faster than the carrier signal~\cite{7335646}. 
When considering correlated channel scenarios, 
the SINR CDF further worsens since a new degree of channel correlation is introduced through the multi-path domain resulting in an ill-conditioned channel. 
At ISDs of 20~m, 50~m and 100~m, 
the difference in 50\%-tile effective SINR between the uncorrelated and correlated scenarios is about 4~dB, 4.85~dB and 6.3~dB, respectively. 
Worth noting that as network becomes denser, 
the difference in 50\%-tile effective SINR between the uncorrelated and correlated channels decreases. 
This is because at very low ISDs, 
the LOS path loss dominates the SINR and hence the impact of spatial correlation and consequently the multi-path fading becomes more trivial. 
Fig.~\ref{fig:Throughput_1} presents a similar comparison in terms of UE throughput CDF of uncorrelated and correlated channels. 
At ISDs of 20~m, 50~m and 100~m
the gain of the 50\%-tile UE throughput from uncorrelated to correlated scenarios is about 1.92x, 1.78x and 1.65x, respectively. 

\begin{figure}[t]
	\centering
	\subfigure[Effective SINR CDF.]{\includegraphics[scale=0.595]{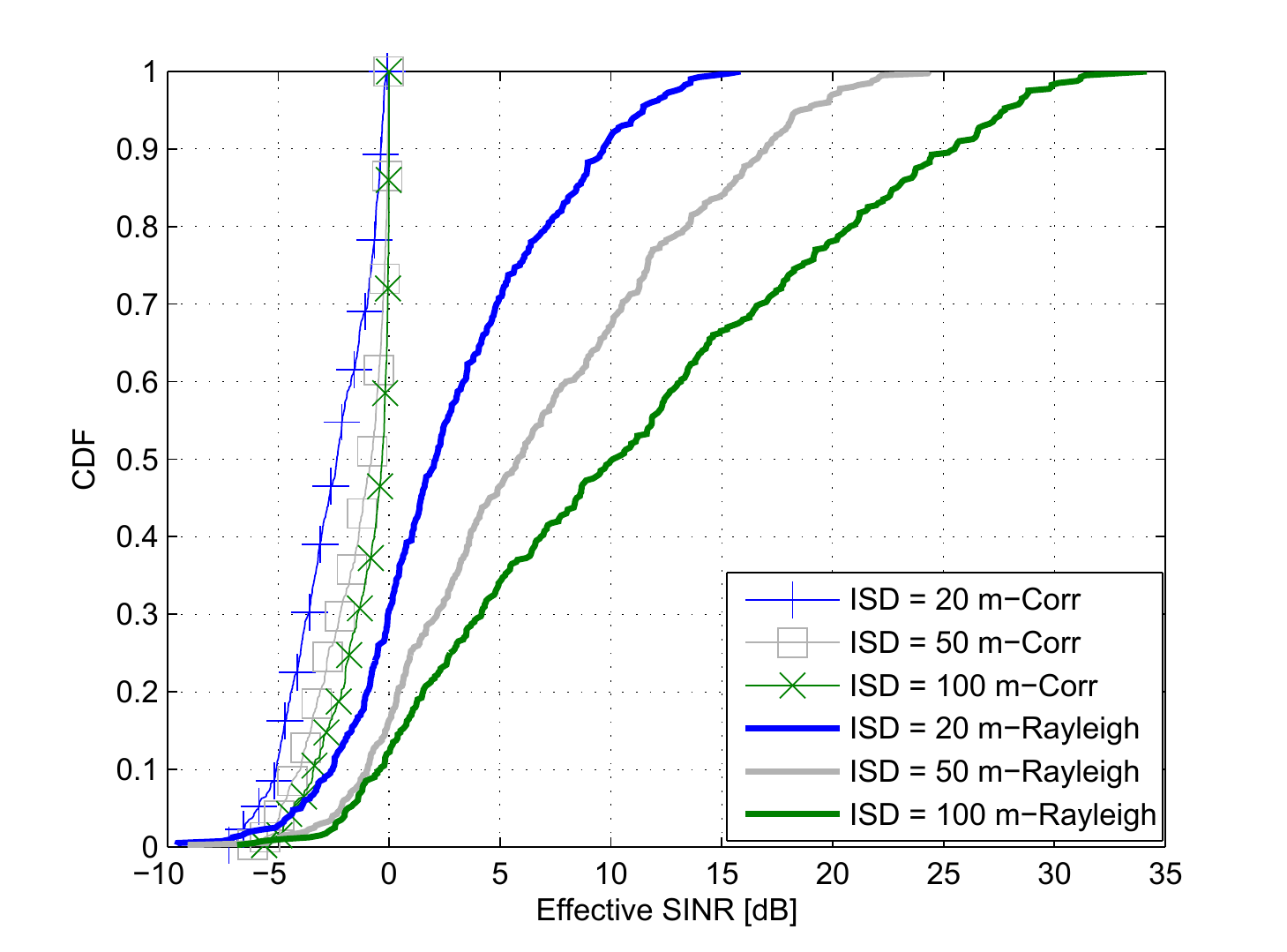}
	\label{fig:SINR_2}}
	\\
	\subfigure[UE throughput CDF.]{\includegraphics[scale=0.595]{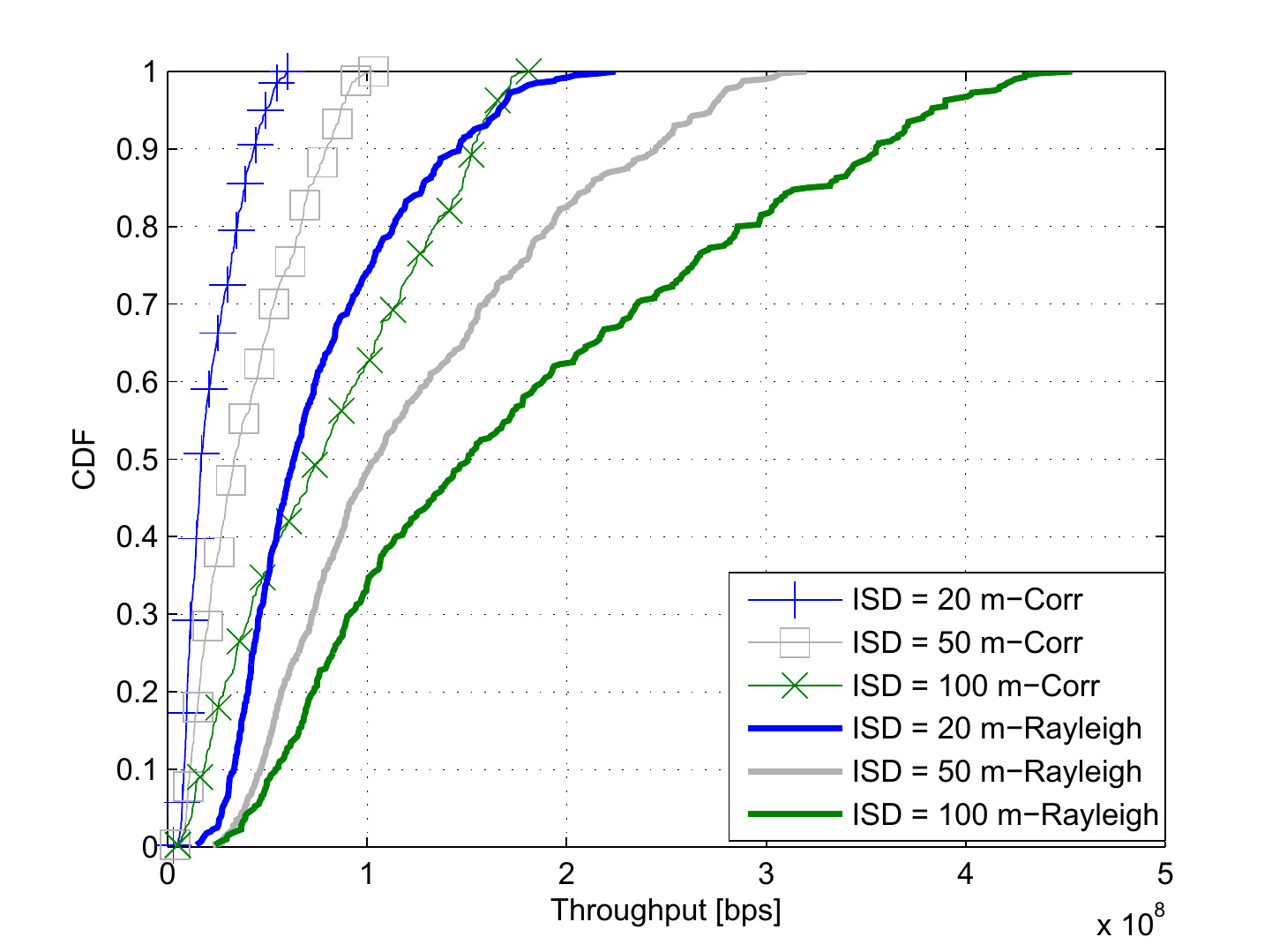}
	\label{fig:Throughput_2}}
	\caption{SINR and throughput CDFs comparison of correlated and Rayleigh channels for a 4x4 MIMO.}
	\label{fig:4x4MIMO_1}
\end{figure}

Figs.~\ref{fig:SINR_2} and~\ref{fig:Throughput_2} show the same performance comparisons for a $4\times4$ MIMO system, 
which can potentially take advantage of upper bound of~4 in terms of the channel degrees of freedom. 
In terms of effective SINR, 
at ISDs of 20~m, 50~m and 100~m,
the difference at 50\%-tile effective SINR between the uncorrelated and correlated systems is about 4.5~dB, 6.8~dB and 10.45~dB, respectively. 
In terms of UE throughput,
at ISDs of 20~m, 50~m and 100~m
the gain of the 50\%-tile UE throughput from uncorrelated to correlated scenarios is about 3.4x, 3.05x and 1.96x, respectively. 
It is worth noting that as the dimension of the MIMO system increases, 
due to the channel higher degrees of freedom, the uncorrelated over correlated gain increases.

\subsection{Performance Enhancement by DPST}

It was shown that the reinforced spatial correlation due to network densification significantly lowers the effective SINR and UE throughput. 
Indeed, it was realised that when channel suffers from larger correlation due to discussed causes, 
the UE receiver is only able to decode the data stream transmitted by the first transmit antenna and fails to decode the data streams sent by the remaining antennas.

In the sequel, DPST is applied to the correlated channel presented in Section~\ref{sec:part_Correlation} to perceive how enhanced diversity among channel pairs (as an outcome of increased ISI due to fractional delay and oversampled receiver) impacts the channel condition number from the receiver perspective. 
Note that in DPST each transmit antenna is subject to a specific fractional delay with respect to its former one, 
which is determined according to Section~\ref{sec:part_Optimization}. 
As discussed, the optimisation seeks to get the Pareto set of fractional deterministic delays corresponding to each transmit antenna that results in the least correlated effective channel. 
In the $2\times2$ MIMO, 
the optimum fractional delay is 0.5 nsec, 
whereas in the $4\times4$ MIMO the optimum Pareto set of fractional delays is [0.25 nsec, 0.45 nsec, 1.5 nsec] applied to second, third and forth transmit antennas. 
Furthermore, the performance of DPST is also compared with respect to the optimistic channel, 
which benefits from channel condition number ($\mathcal{K}$) equals to~1. 
This refers to the optimal channel condition where all the channels are orthogonal, 
and hence the UE throughput can be enhanced by the minimum number of transmit/receiver antennas.

\begin{figure}
	\centering
	\subfigure[Effective SINR CDF.]{\includegraphics[scale=0.595]{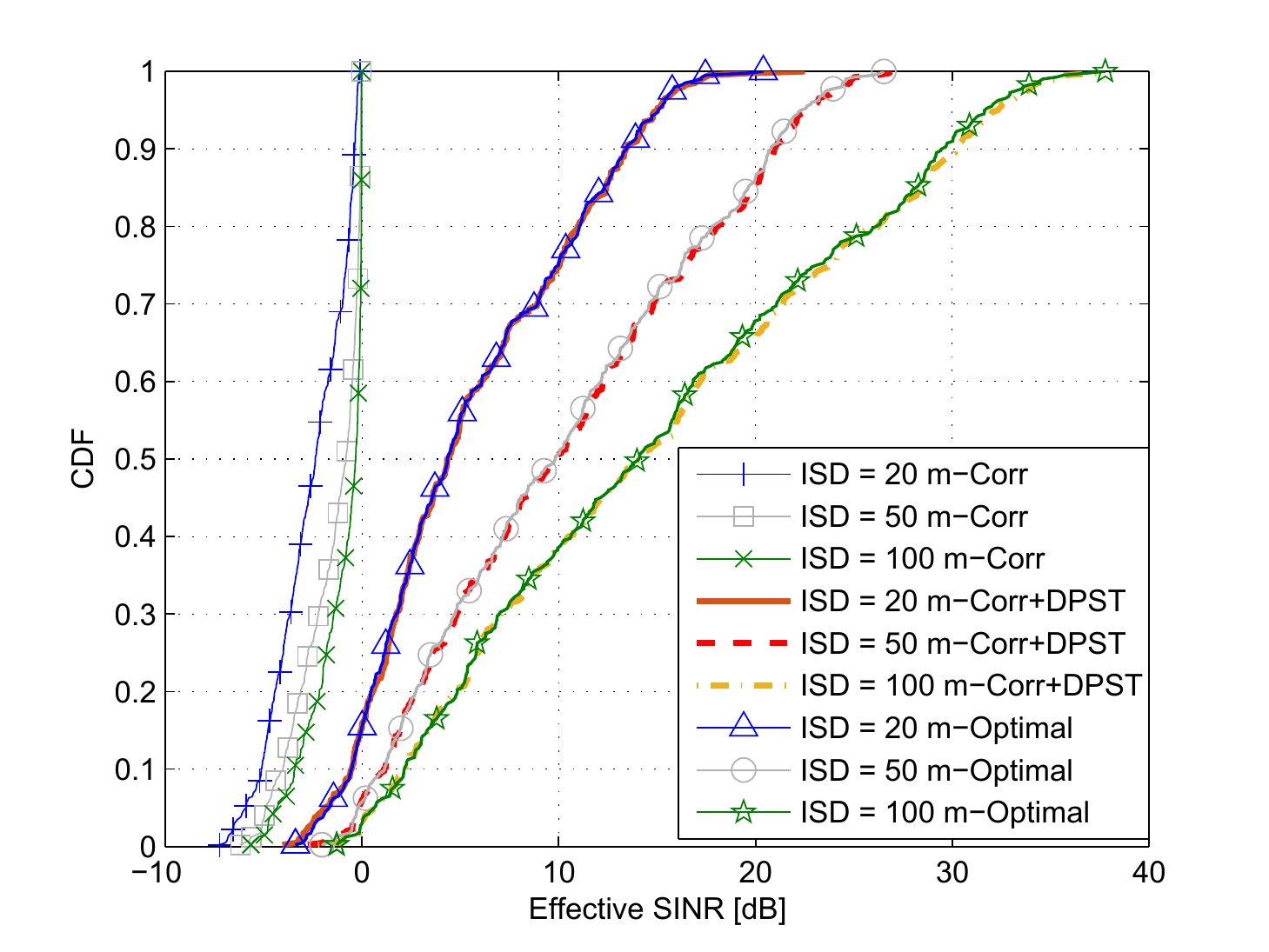}
	\label{fig:SINR_3}}
	\\
	\subfigure[UE throughput CDF.]{\includegraphics[scale=0.595]{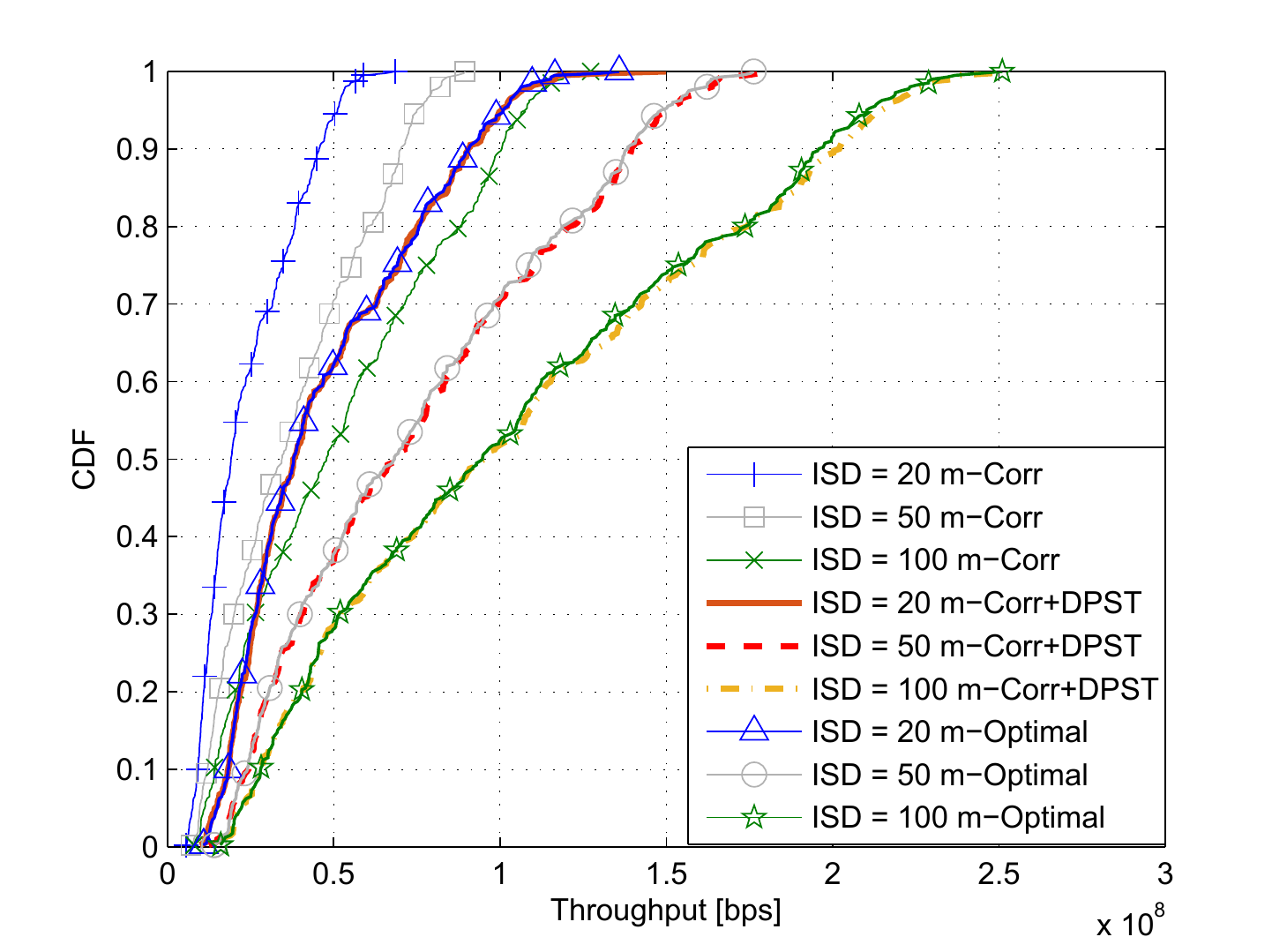}
	\label{fig:Throughput_3}}
	\caption{SINR and throughput CDFs comparison of correlated, DPST and optimum channel conditions for a 2x2 MIMO.}
	\label{fig:2x2MIMO_2}
\end{figure}

Figs.~\ref{fig:SINR_3} and~\ref{fig:Throughput_3} compare the effective SINR and UE throughput CDFs for a $2\times2$ MIMO system where DPST is applied to the correlated channel. 
Results show that at all tested ISDs, 
DPST can significantly improve the effective SINR and UE throughput with respect to the case of no DPST, 
offering a very close to optimal performance. The attained 50\%-tile SINR improvement by DPST with respect to correlated scenario -- where no DPST is applied -- at ISDs of 20~m, 50~m and 100~m is about 6.7~dB, 10.2~dB and 13.6~dB, respectively. At all respective ISDs, DPST performance is only about 0.14~dB away from the optimum performance. 
 
\begin{figure}[t]
\centering
\includegraphics[scale=0.595]{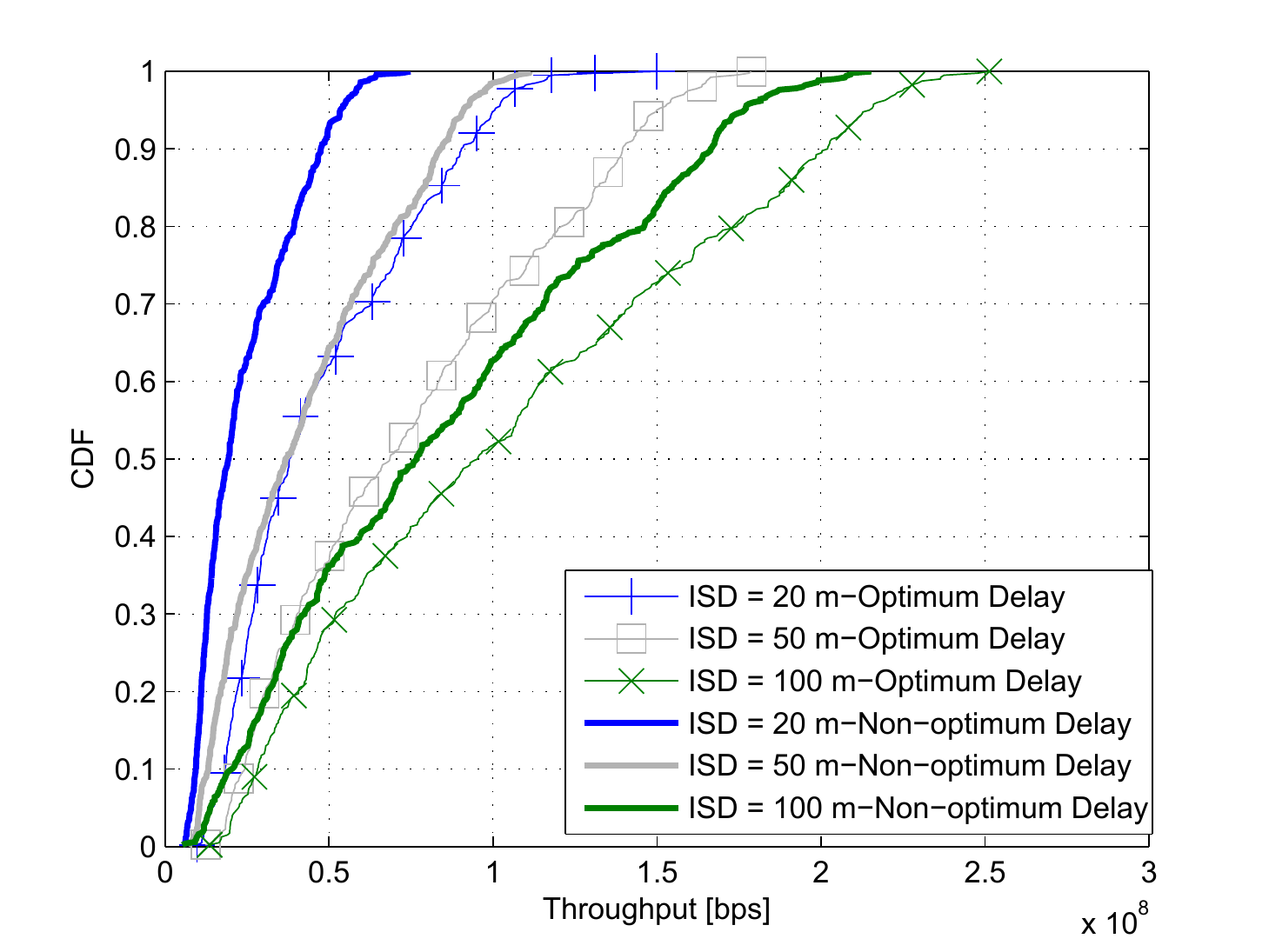}
\caption{Impact of optimum and non-optimum delays on UE throughput CDF.}
\label{fig:Optimum_delay}
\vspace{-0.3cm}
\end{figure}

Note that this significant performance enhancement by DPST is because through the optimisation of deterministic delay, 
DPST is able to lower the condition number of the virtual channel as close as possible to 1. 
Accordingly, DPST is also able to boost the UE throughput by almost 1.93x at all respective ISDs, 
which is less than 1.03 away from optimum performance. 
This implies that even at very low ISDs, 
where the channel struggles with a high degree of spatial correlation, 
DPST remarkably enhances the UE throughput through lowering the channel condition number. 
Note that the choice of optimum delay plays a key role in DPST performance. 
Fig.~\ref{fig:Optimum_delay} shows that in a $2\times2$ MIMO system and at ISDs of 20~m, 50~m and 100~m, 
the optimum delay boosts the UE throughput by 1.95x, 1.86x, 1.4x with respect to non-optimum delay of 1.3 nsec.

\begin{figure}
	\centering
	\subfigure[Effective SINR CDF.]{\includegraphics[scale=0.595]{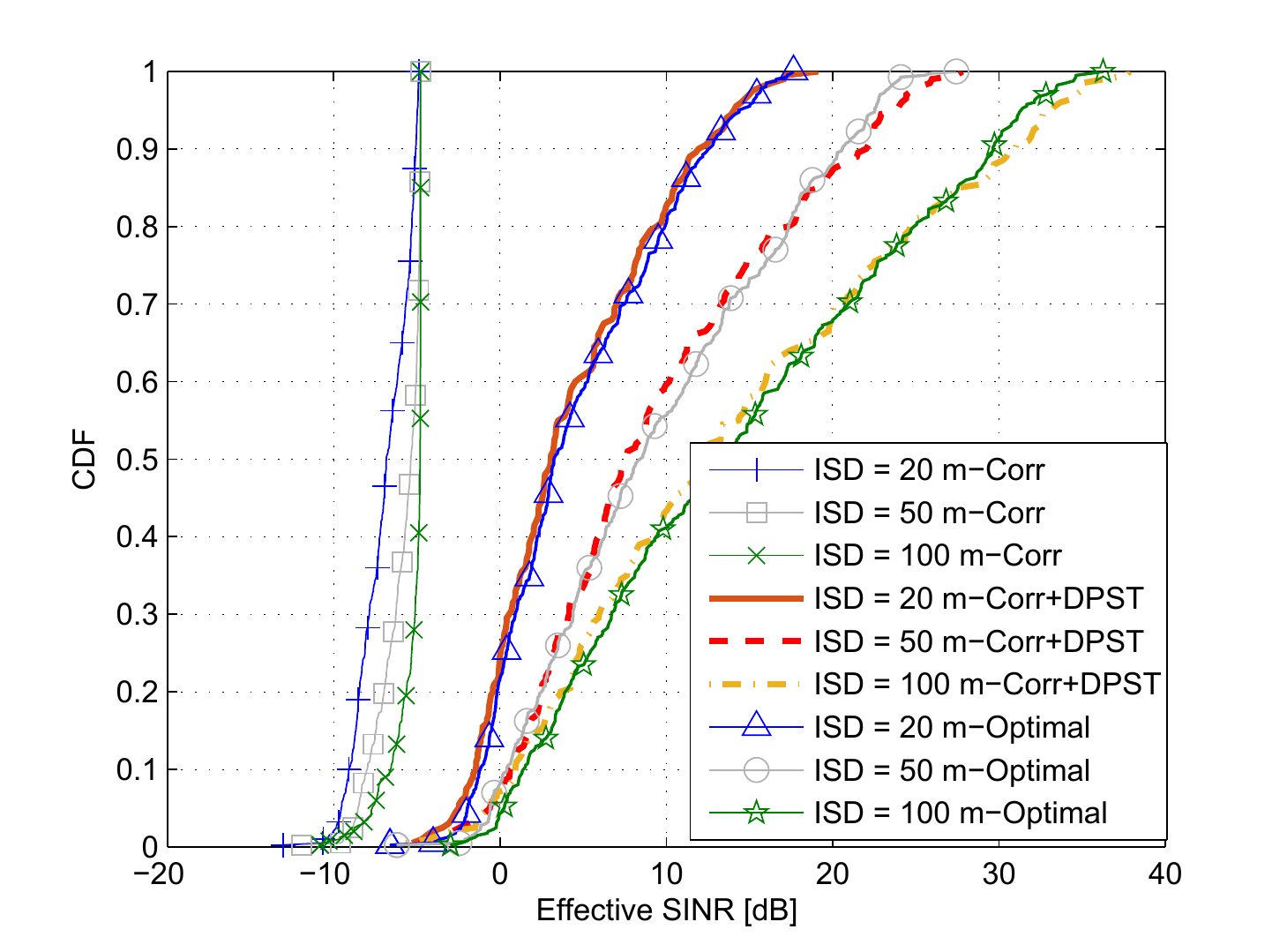}
	\label{fig:SINR_4}}
	\\
	\subfigure[UE throughput CDF.]{\includegraphics[scale=0.595]{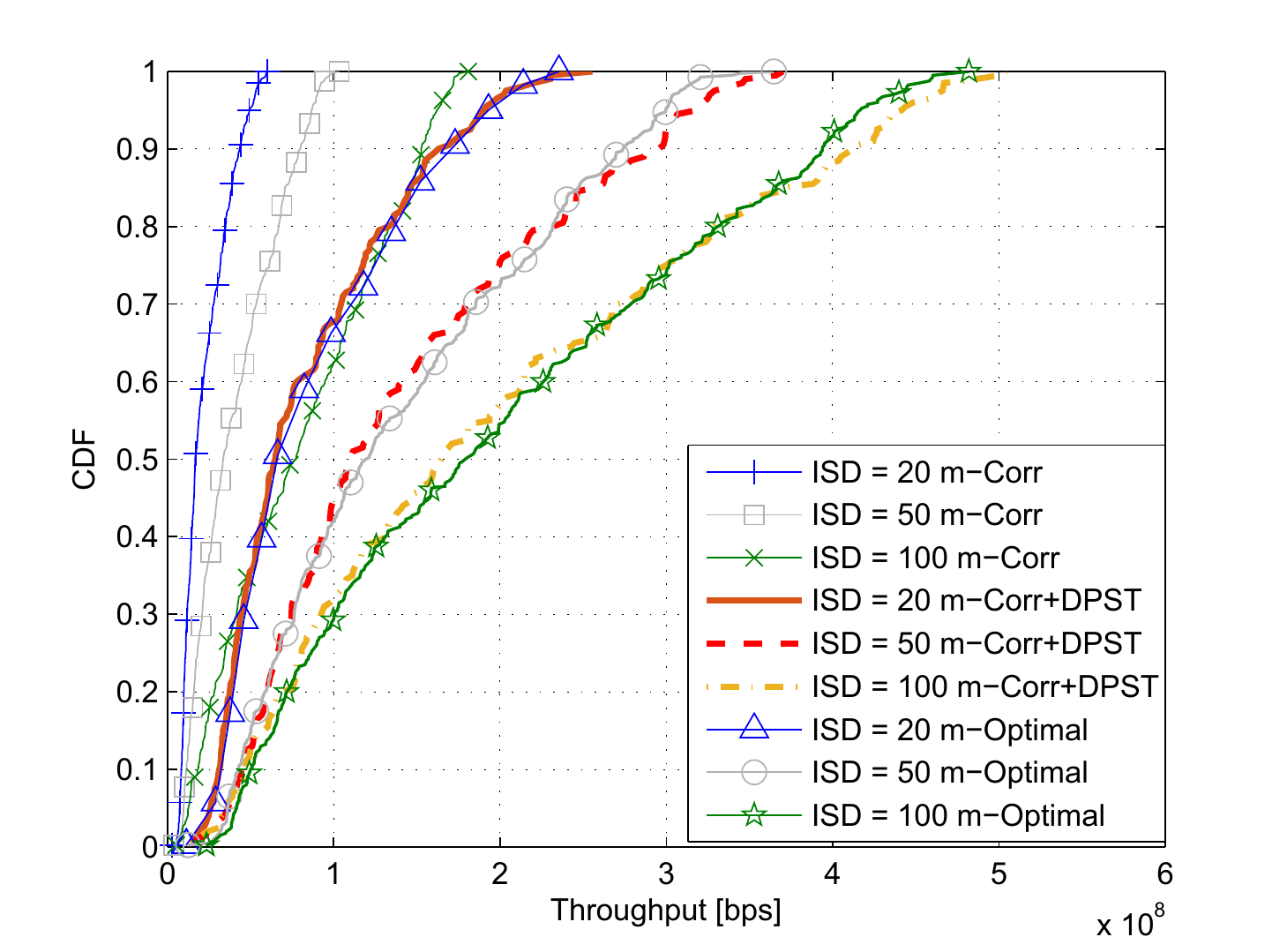}
	\label{fig:Throughput_4}}
	\caption{SINR and throughput CDFs comparison of correlated, DPST and optimum channel conditions for a 4x4 MIMO.}
	\label{fig:4x4MIMO_2}
\end{figure}

Similarly, Figs.~\ref{fig:SINR_4} and~\ref{fig:Throughput_4} compare the performances of a $4\times4$ MIMO system. 
Likewise, DPST enhances the effective SINR and UE throughput. In more detail, at ISDs of 20~m, 50~m and 100~m, 
DPST enhances the 50\%-tile wideband SINR with respect to the correlated channel by 9.6~dB, 12.6~dB and 14.8~dB, respectively. 
However, the extent of improvement with respect to optimum scenario ($\mathcal{K} = 1$) differs from that of the $2\times2$ MIMO system. Performance from the optimum is about 0.15~dB, 0.8~dB and 1~dB away at respective ISDs. 
In terms of the UE throughput, 
it can perceived that UE throughput is enhanced by about 3.76x when DPST is exploited and the difference between DPST and optimum scenarios at all ISDs is about 1.13x.

\begin{table}[tbph]
\centering
\renewcommand{\arraystretch}{1.4}
\caption{Performance comparison under different channel condition}
\label{tab:channel_cond}
\scalebox{0.85}{
\centering
\begin{tabular}{|>{\centering\arraybackslash}p{2.9cm}|>{\centering\arraybackslash}p{1.9cm}|>{\centering\arraybackslash}p{3.6cm}|}
\hline \textbf{Channel Status} & \textbf{Channel Rank} & \textbf{Channel Condition Number} \\ \hline
\hline
LOS channel (Default) & 1 & $\infty$     \\ \hline
2 $\times$ 2 Rayleigh & 2 & 4.45     \\ \hline
2 $\times$ 2 DPST & 2 & 1.31     \\ \hline
2 $\times$ 2 Optimum  & 2  & 1 \\ \hline
4 $\times$ 4 Rayleigh & 4 & 9.35     \\ \hline
4 $\times$ 4 DPST & 4 & 1.42     \\ \hline
4 $\times$ 4 Optimum & 4 & 1     \\ \hline
\end{tabular}}
\vspace{-0.2cm}
\end{table}

Table \ref{tab:channel_cond} summarizes the rank and condition number of different channel scenarios. 
Beware that DPST refers to the virtual channel attained after applying DPST with optimum deterministic fractional delay to the correlated channel. 
Also, it must be stated that the condition number is obtained over an average of 10000 different realisations of the channel type. 
As shown in table \ref{tab:channel_cond}, 
the system performance when the DPST is used outperforms that of the Rayleigh channel. 
While Rayleigh channel offers a well conditioned channel, 
it is still far from the optimal channel (condition numbers of 4.45 and 9.35 for $2\times2$ and $4\times4$ MIMO systems, respectively). 
In contrast, the combined effects of the {\rm{i)}} optimised deterministic fractional delay interpolated with sinc pulse shaping and the {\rm{ii)}} the oversampled receiver helps DPST to forcefully converge to a virtual channel with a condition number of close to 1. Note that this capacity boost will also enhance the energy efficiency of the network, since it allows to transmit more bits at the same energy consumption rate.

\section{Conclusion}
\label{sec:part_Conclusion}
 
In this paper, we discussed that in ultra-dense small cell networks, 
the presence of spatial channel correlation is an obstacle to achieve spatial multiplexing gain. 
We presented a model for the channel correlation, 
which captures the effects of both UE-BS distance and antenna spacing as the main sources of spatial correlation. 
We proposed a new technique referred to as diversity pulse shaped transmission (DPST), 
which modulates the transmission of adjacent antennas with distinct interpolated shaped pulses that are offset by a deterministic delay 
(which is a fraction of symbol period). 
At the receiver side, DPST takes advantage of fractionally spaced equaliser (FSE) that operates in the oversampled domain. 
The combined effect of DPST and FSE is able to significantly enhance the diversity among the channel pairs, 
and thus generates a virtual channel with reduced correlation among its pairs from the receiver perspective. 
We studied the performance of the proposed technique under MMSE criteria, 
and compared it with existing $2\times2$ and $4\times4$ MIMO cellular configurations. 
We showed that DPST can enhance the UE throughput by 1.93x and 3.76x in $2 \times 2$ and $4 \times 4$ MIMO systems, respectively. As part of future research, we will analyse the energy efficiency of ultra-dense small cell networks while applying DPST.

\appendix
\label{appendix1}

We consider a MIMO system with $N_{t}$ transmit and $N_{r}$ receive antennas with corresponding antenna spacings of $d_{t}$ and $d_{r}$, respectively.
It was discussed in section (\ref{sec:part_Correlation}) that the correlated Rician channel can be decomposed to LOS and NLOS matrices. 
Considering the correlated NLOS MIMO channel, 
the degree of spatial correlation between any two transmit-receive pair in a $N_{r}\times N_{t}$ MIMO is computed as

\vspace{-0.3cm}
\begin{equation}
\rho_{ij,pq} = \frac{E[h_{i,j}h_{p,q}^{*}]}{\sqrt{E[h_{i,j}h_{i,j}^{*}]E[h_{p,q}h_{p,q}^{*}]}} \quad
\begin{cases}
\textit{i,p} = 1,2,...,N_{r} \\ \textit{j,q} = 1,2,...,N_{t} 
\end{cases}
\label{eq:correlation_equation}
\end{equation}
where $h_{i,j}$ is the channel from $j^{th}$ transmit antenna to $i^{th}$ receive antenna and $\textit{E}$ refers to the expectation operation.

For sake of clarity, we only consider a $2\times2$ MIMO system in this paper. 
The channel matrix in a 2$\times$2 MIMO is given by

\[
\textbf{H} = 
\left[\bea{cc}
h_{1,1}(t) & h_{1,2}(t)
\\ h_{2,1}(t) & h_{2,2}(t)
\ena\right]
\]

From (\ref{eq:correlation_equation}), 
the spatial correlation coefficient between channel pairs, $h_{1,1}$ and $h_{2,2}$ is
\vspace{-0.2cm}
\begin{equation*}
\rho_{11,22} = \frac{E[h_{1,1}h_{2,2}^{*}]}{\sqrt{E[h_{1,1}h_{1,1}^{*}]E[h_{2,2}h_{2,2}^{*}]}}
\end{equation*}
\vspace{-0.4cm}

To derive the correlation coefficient as a function of antenna spacing, 
we model the channel gains as the ratio of received output voltage $\textbf{v}_{out}$ to the transmitted input voltage $\textbf{v}_{in}$. 
Noteworthy that the transmitted input voltage is assumed identical for all transmit antennas~\cite{Correlation_Sing}.

\[
\bea{ccc}
\left[\bea{c}
v_{out,1}(t) \\ v_{out,2}(t)
\ena\right]
& = &
\left[\bea{cc}
h_{1,1}(t) & h_{1,2}(t) 
\\ h_{2,1}(t) & h_{2,2}(t)
\ena\right]
\ast
\left[\bea{c}
v_{in}(t) \\ v_{in}(t)
\ena\right]
\vspace{0.15cm}
\\
\left[\bea{c}
v_{out,1}(t) \\ v_{out,2}(t)
\ena\right]
& = &
\left[\bea{cc}
v_{out,1,1}(t) + v_{out,1,2}(t) 
\\ v_{out,2,1}(t) + v_{out,2,2}(t)
\ena\right]
\vspace{0.2cm}
\\
\bv_{out}(t) & = & \bH(t) \bv_{in}(t)
\ena
\]
where $v_{out,i,j}$ refers to the signal received by $i^{th}$ receive antenna from $j^{th}$ transmit antenna, respectively. 
The correlation coefficient between $h_{1,1}$ and $h_{2,2}$ is therefore written as 

\begin{equation}
\vspace{-0.3cm}
\rho_{11,22} = \frac{E[v_{out,1,1}v_{out,2,2}^{*}]}{\sqrt{E[v_{out,1,1}v_{out,1,1}^{*}]E[v_{out,2,2}v_{out,2,2}^{*}]}}
\label{eq:Correlation_voltage}
\vspace{0.2cm}
\end{equation}

For simplicity, we only compute the numerator of (\ref{eq:Correlation_voltage}) reminding that its denominator can be derived in the same way. 
Considering that electric voltage can be computed as the integral of electric field over the path taken, 
the numerator of (\ref{eq:Correlation_voltage}) is written as

\vspace{-0.4cm}
\begin{equation*}
\begin{split}
	& E[v_{out,1,1} v_{out,2,2}^{*}] = E[[\frac{-1}{I}\int_{0}^{2\pi}\int_{0}^{l}I(z)(\int_{0}^{2\pi} e_{1}(\phi) d\phi)\\
	& E_{1}(\varphi) dz d\varphi]
	[\frac{-1}{I}\int_{0}^{2\pi}\int_{0}^{l}I(z)(\int_{0}^{2\pi}
	 e_{2}(\phi) d\phi) E_{2}(\varphi) dz d\varphi]^{*}] \\
	 & = (\frac{-1}{I^{2}})[\int_{0}^{l}\int_{0}^{l}I(z)I^{*}(z) dz
	 	E[(\int_{0}^{2\pi} e_{1}(\phi) d\phi \int_{0}^{2\pi} e_{2}^{*}(\phi) d\phi^{*}) \\
	 	& (\int_{0}^{2\pi} E_{1}(\varphi) d\varphi \int_{0}^{2\pi} E_{2}({\varphi^{'}}) d\varphi^{'})] d{z^{*}}] 
\end{split}
\label{eq:electric_field1}
\end{equation*}
where \textit{I} is the induced current in the antennas, 
$E_{1}$ and $E_{2}$ refer to incident fields at the receive antennas 
and $e_{1}$ and $e_{2}$ are the far fields generated by the transmit antennas. 
$\varphi$ and $\phi$ also denote the angle of departure (AoD) from transmit antennas and angle of arrival (AoA) at receive antennas, respectively. 
Due to different spatial locations of the antennas, 
$e_{2}(\phi)$ and $E_{2}(\varphi)$ are modelled as $e_{0} e^{j d_{t} cos(\phi)}$ and $E_{0} e^{j d_{r} cos(\varphi)}$, respectively, 
where $e_{1}(\phi)= e_{0}$ and $E_{1}(\varphi)= E_{0}$. Note that $e_{0}$ and $E_{0}$ follow Rayleigh distribution.

\begin{equation*}
\begin{split}
& E[v_{out,1,1} v_{out,2,2}^{*}] = (\frac{-1}{I^{2}})[\int_{0}^{l}\int_{0}^{l}I(z)I^{*}(z) dz d{z^{*}}]
E[(\int_{0}^{2\pi} \\
& e_{0} \int_{0}^{2\pi} e_{0}e^{j d_{t} cos(\phi^{*})} d\phi^{*})(\int_{0}^{2\pi}E_{0} \int_{0}^{2\pi} E_{0}e^{j d_{r} cos(\varphi^{'})} d\varphi^{'})] \\ 
& =(\frac{-1}{I^{2}})[\int_{0}^{l}\int_{0}^{l}I(z)I^{*}(z) dz d{z^{*}}] E[\int_{0}^{2\pi} \int_{0}^{2\pi} e_{0}^2 E_{0}^2 e^{j d_{t} cos(\phi^{*})} \\ 
& d\phi ^{*} e^{j d_{r} cos(\varphi^{'})} d\varphi^{'}]
\end{split}
\label{eq:electric_field}
\end{equation*}

Intuitively, the two inner integrals represent the auto-correlation functions of the two Rayleigh distributed variables, $e_{0}$ and $E_{0}$. 
Provided that the auto-correlation function of Rayleigh distribution can be modelled by Bessel function, 
it follows that
\vspace{-0.2cm}
\begin{equation*}
\begin{split}
	&E[v_{out,1,1} v_{out,2,2}^{*}] = (\frac{-1}{I^{2}})[\int_{0}^{l}\int_{0}^{l}I(z)I^{*}(z) dz d{z^{*}}] \\ &J_{0}(2\pi d_{t}) J_{0}(2\pi d_{r}) 
\end{split}
\end{equation*}
where $J_{0}$ is the zero order first kind Bessel function. 
The denominator of (\ref{eq:Correlation_voltage}) can be expanded likewise, 
and, therefore, it can be shown that the correlation coefficient between any two channel pairs in a MIMO system with $N_{t}$ and $N_{r}$ transmit and receive antennas spaced by $d_{t}$ and $d_{r}$ can be computed as

\vspace{-0.3cm}
\begin{equation*}
\rho_{ij,pq} = J_{0}(2\pi d_{t}|q-j|) J_{0}(2\pi d_{r}|p-i|) \quad
\begin{cases}
\textit{i,p} = 1,2,...,N_{r} \\ \textit{j,q} = 1,2,...,N_{t} 
\end{cases}
\label{eq:Spatial_correlation}
\end{equation*}
where $\mid \mid$ denotes the absolute operation. 
Having computed the correlation coefficient between any two channel pairs, 
the transmit and receive correlation matrices respectively denoted by $\textbf{R}_{T}$ and $\textbf{R}_{R}$ are derived, 
which take into account the impact of antenna spacing. 
This is used to determine the correlated channel model in (\ref{eq:Full_correlated_channel}) that considers both the UE-BS distance and antenna spacing.

\bibliographystyle{ieeetr}
\bibliography{references}

\begin{thebibliography}{10}

\bibitem{2015Lopez}
D.~L\'opez-P\'erez, M.~Ding, H.~Claussen, and A.~H. Jafari, ``{Towards 1
  {G}bps/{UE} in cellular systems: understanding ultra-dense small cell
  deployment},'' in {\em IEEE Commun. Surveys Tuts.}, Jun. 2015.

\bibitem{Eurasip}
A.~H. Jafari, D.~L\'opez-P\'erez, H.~Song, H.~Claussen, L.~Ho, and J.~Zhang,
  ``Small cell backhaul: challenges and prospective solutions,'' {\em {EURASIP}
  J. on Wireless Commun. and Netw.}, vol.~2015, no.~1, pp.~1--18, 2015.

\bibitem{fronthaul}
H.~Zhang, Y.~Dong, J.~Cheng, M.~J. Hossain, and V.~C.~M. Leung, ``Fronthauling
  for {5G LTE-U} ultra dense cloud small cell networks,'' {\em \rm {[}Online{]}
  Available: http://arxiv.org/abs/1607.07015}, Sep. 2016.

\bibitem{Nokia_UDN}
{Nokia}, ``{Ultra Dense Network (UDN)},'' 2016.

\bibitem{1532224}
J.~Akhtar and D.~Gesbert, ``Spatial multiplexing over correlated {MIMO}
  channels with a closed-form precoder,'' {\em IEEE Trans. Wireless Commun.},
  vol.~4, pp.~2400--2409, Sep. 2005.

\bibitem{LTE}
F.~Khan, {\em {{LTE} for {4G} Mobile Broadband Air Interface Technologies and
  Performance}}.
\newblock University Cambridge Press, 2009.

\bibitem{Jafari-SPAWC}
A.~H. Jafari, V.~Venkateswaran, D.~L\'opez-P\'erez, and J.~Zhang, ``Pulse
  shaping diversity to enhance throughput in ultra-dense small cell networks,''
  in {\em Proc. of IEEE International Workshop on SPAWC}, pp.~1--5, Jul. 2016.

\bibitem{489269}
J.~Treichler, I.~Fijalkow, and C.~Johnson, ``Fractionally spaced equalizers,''
  {\em IEEE Signal Process. Mag.}, vol.~13, pp.~65--81, May 1996.

\bibitem{1203167}
M.~Ivrlc, W.~Utschick, and J.~Nossek, ``Fading correlations in wireless {MIMO}
  communicatin systems,'' {\em IEEE J. Sel. Areas Commun.}, vol.~21,
  pp.~819--828, Jun. 2003.

\bibitem{1459054}
A.~M. Tulino, A.~Lozano, and S.~Verdu, ``Impact of antenna correlation on the
  capacity of multiantenna channels,'' {\em IEEE Trans. Inf. Theory}, vol.~51,
  pp.~2491--2509, Jul. 2005.

\bibitem{892194}
D.~Chizhik, F.~Rashid-Farrokhi, J.~Ling, and A.~Lozano, ``Effect of antenna
  separation on the capacity of {BLAST} in correlated channels,'' {\em IEEE
  Commun. Lett.}, vol.~4, pp.~337--339, Nov. 2000.

\bibitem{2015Jafari}
A.~H. Jafari, D.~L\'opez-P\'erez, M.~Ding, and J.~Zhang, ``{Study on Scheduling
  Techniques for Ultra Dense Small Cell Networks},'' in {\em Proc. of IEEE VTC
  Fall}, pp.~1--6, Sep. 2015.

\bibitem{1683382}
C.~Oestges, ``Validity of the kronecker model for {MIMO} correlated channels,''
  in {\em Proc. of IEEE VTC Spring}, vol.~6, pp.~2818--2822, May 2006.

\bibitem{Matrix}
R.~B. Bapat, S.~J. Kirkland, K.~M. Prasad, and S.~Puntanen, {\em {Combinatorial
  Matrix Theory and Generalized Inverses of Matrices}}.
\newblock Springer, 2013.

\bibitem{5956339}
J.~Shen, Y.~Oda, T.~Furuno, T.~Maruyama, and T.~Ohya, ``A novel approach for
  capacity improvement of 2x2 {MIMO} in {LOS} channel using reflectarray,'' in
  {\em Proc. of IEEE VTC Spring}, pp.~1--5, May 2011.

\bibitem{6479673}
J.~Anderson, F.~Rusek, and V.~Owall, ``{F}aster-than-{N}yquist signaling,''
  {\em Proc. of the IEEE}, vol.~101, pp.~1817--1830, Aug. 2013.

\bibitem{4777625}
F.~Rusek and J.~Anderson, ``Constrained capacities for {F}aster-than-{N}yquist
  signaling,'' {\em IEEE Trans. Inf. Theory}, vol.~55, pp.~764--775, Feb. 2009.

\bibitem{1231648}
A.~Liveris and C.~Georghiades, ``Exploiting faster-than-{N}yquist signaling,''
  {\em IEEE Trans. Commun.}, vol.~51, pp.~1502--1511, Sep. 2003.

\bibitem{6620766}
A.~Sheikholeslami, D.~Goeckel, and H.~Pishro-Nik, ``Artificial intersymbol
  interference (isi) to exploit receiver imperfections for secrecy,'' in {\em
  Proc. of IEEE ISIT}, pp.~2950--2954, Jul. 2013.

\bibitem{5706377}
V.~Venkateswaran and A.-J. van~der Veen, ``Multichannel {S}igma{D}elta {ADC}s
  with integrated feedback beamformers to cancel interfering communication
  signals,'' {\em IEEE Trans. Signal Process.}, vol.~59, pp.~2211--2222, May
  2011.

\bibitem{4656965}
S.~Plass, A.~Dammann, and S.~Sand, ``An overview of cyclic delay diversity and
  its applications,'' in {\em Proc. of IEEE VTC Fall}, pp.~1--5, Sep. 2008.

\bibitem{989781}
W.~Lee, ``The most spectrum-efficient duplexing system: {CDD},'' {\em IEEE
  Commun. Mag.}, vol.~40, pp.~163--166, Mar. 2002.

\bibitem{4109390}
S.~Plass and A.~Dammann, ``Cellular cyclic delay diversity for next generation
  mobile systems,'' in {\em Proc. of IEEE VTC Fall}, pp.~1--5, Sep. 2006.

\bibitem{5426062}
G.~Bauch and T.~Abe, ``On the parameter choice for cyclic delay diversity based
  precoding with spatial multiplexing,'' in {\em Proc. of IEEE Globecom},
  pp.~1--6, Nov. 2009.

\bibitem{4212739}
M.~I. Rahman, S.~S. Das, E.~de~Carvalho, and R.~Prasad, ``Spatial multiplexing
  in {OFDM} systems with cyclic delay diversity,'' in {\em Proc. of IEEE VTC
  Spring}, pp.~1491--1495, Apr. 2007.

\bibitem{5768230}
C.~Ma, ``Mitigation of transmit correlation for {MIMO} spatial multiplexing
  through phase rotation precoding,'' in {\em Proc. of CECNet}, pp.~1398--1401,
  Apr. 2011.

\bibitem{1192168}
D.~Gesbert, M.~Shafi, D.-S. Shiu, P.~Smith, and A.~Naguib, ``From theory to
  practice: an overview of {MIMO} space-time coded wireless systems,'' {\em
  IEEE J. Sel. Areas Commun.}, vol.~21, pp.~281--302, Apr. 2003.

\bibitem{7270991}
H.~Pervaiz, L.~Musavian, Q.~Ni, and Z.~Ding, ``Energy and spectrum efficient
  transmission techniques under {Q}o{S} constraints toward green heterogeneous
  networks,'' {\em IEEE Access}, vol.~3, pp.~1655--1671, 2015.

\bibitem{7496967}
Z.~Song, Q.~Ni, K.~Navaie, S.~Hou, S.~Wu, and X.~Sun, ``On the spectral-energy
  efficiency and rate fairness tradeoff in relay-aided cooperative {OFDMA}
  systems,'' {\em IEEE Trans. Wireless Commun.}, vol.~15, pp.~6342--6355, Sep.
  2016.

\bibitem{3gpp}
``{3{GPP} TSG RAN, TR 25.996 v10.0.0, "Spatial Channel Model for Multiple Input
  Multiple Output (MIMO) simulations (release 10)},'' Mar. 2011.

\bibitem{7335646}
M.~Ding, P.~Wang, D.~L\'opez-P\'erez, G.~Mao, and Z.~Lin, ``Performance impact
  of {L}o{S} and {NL}o{S} transmissions in dense cellular networks,'' {\em IEEE
  Trans. Wireless Commun.}, vol.~15, pp.~2365--2380, Mar. 2016.

\bibitem{Correlation_Sing}
H.~T. Hui, {\em {Multiple Antennas for {MIMO} Communciations-Channel
  Correlation}}.
\newblock National University of Singapore.

\end{thebibliography}

\end{document}